\documentstyle[12pt]{article} 
\newcommand{\rf}[1]{(\ref{#1})}
\newcommand{\bea}{\begin{eqnarray}}
\newcommand{\eea}{\end{eqnarray}}

\newcommand{\g}{\gamma}
\renewcommand{\l}{\lambda}
\renewcommand{\L}{\Lambda}
\renewcommand{\b}{\beta}
\renewcommand{\a}{\alpha}

\newcommand{\m}{\mu}

\newcommand{\th}{\theta}
\newcommand{\ep}{\varepsilon}
\newcommand{\EP}{{\cal E}}

\newcommand{\sg}{\sigma}
\newcommand{\k}{\kappa}

\newcommand{\tr}{{\rm Tr}\,}

\newcommand{\sdet}{{\rm sdet}\,}

\newcommand{\prt}{\partial}

\newcommand{\half}{\frac{1}{2}}

\newcommand{\quarter}{\frac{1}{4}}

\newcommand{\cL}{{\cal L}}

\newcommand{\Htr}{{\rm Htr}\,}

\newcommand{\dd}{{\rm d}}
\newcommand{\DD}{{\cal D}}
\newcommand{\bDD}{{\rm D}}

\def\void{}
\def\labelmark{}

\newenvironment{formula}[1]{\def\labelname{#1}
\ifx\void\labelname\def\junk{\begin{displaymath}}
\else\def\junk{\begin{equation}\label{\labelname}}\fi\junk}%
{\ifx\void\labelname\def\junk{\end{displaymath}}
\else\def\junk{\end{equation}}\fi\junk\labelmark\def\labelname{}}

{\ifx\void\labelname\def\junk{\end{array}\end{displaymath}}
\else\def\junk{\end{array}\right.\end{equation}}
\fi\junk\labelmark\def\labelname{}\def\junk{}
}

\newcommand{\beq}{\begin{formula}}
\newcommand{\eeq}{\end{formula}}
\newcommand{\beqv}{\begin{formula}{}}

\newcommand{\AuxA}{{\Phi}}
\newcommand{\auxA}{{\phi}}
\newcommand{\auyA}{{\kappa}}
\newcommand{\auzA}{{G}}

\newcommand{\tAA}{{\tilde \Psi}}
\newcommand{\tA}{{\tilde A}}
\newcommand{\Ay}{{\psi}}
\newcommand{\hAy}{{\hat \psi}}
\newcommand{\Az}{{X}}
\newcommand{\tAz}{{X'}}

\newcommand{\tBB}{{\tilde \Pi}}
\newcommand{\tB}{{\tilde B}}
\newcommand{\By}{{\rho}}
\newcommand{\hBy}{{\hat \rho}}
\newcommand{\Bz}{{Y}}
\newcommand{\tBz}{{Y'}}

\newcommand{\tCC}{{\tilde \Lambda}}
\newcommand{\tC}{{\tilde C}}
\newcommand{\Cy}{{\pi}}
\newcommand{\Cz}{{H}}

\newcommand{\tV}{{V'}}
\newcommand{\tv}{{v'}}
\newcommand{\tmu}{{\mu'}}
\newcommand{\tM}{{M'}}

\newcommand{\tW}{{\tilde W}}
\newcommand{\tw}{{\tilde w}}
\newcommand{\tf}{{f'}}
\newcommand{\htau}{{\hat \tau}}

\newcommand{\UvUa}{${\rm U}(1)_{\rm V} \!\times\! {\rm U}(1)_{\rm A}$}

\begin{document}
\topmargin 0pt
\oddsidemargin 5mm                              
\headheight 0pt
\headsep 0pt
\topskip 9mm

\hfill  TIT/HEP--494

\hfill  Feb. 2003


\begin{center}

  \vspace{24pt}
  {\large \bf 
     The bosonic string and superstring models \\
     in $26+2$ and $10+2$ dimensional space--time, \\
     and the generalized Chern-Simons action
  }

  \vspace{24pt}

  {\sl Yoshiyuki Watabiki}

  \vspace{6pt}

  Department of Physics\\
  Tokyo Institute of Technology\\
  Oh-okayama, Meguro, Tokyo 152, Japan\\

\end{center}

\vspace{24pt}

\begin{center}
 {\bf Abstract}
\end{center}
\vspace{12pt}

\noindent


We have covariantized the Lagrangians of the {\UvUa} models, 
which have {\UvUa} gauge symmetry in two dimensions, 
and studied their symmetric structures. 
The special property of the {\UvUa} models is the fact that 
all these models have an extra time coordinate 
in the target space--time. 
The {\UvUa} models coupled to two-dimensional gravity
are string models 
in $26+2$ dimensional target space--time 
for bosonic string and 
in $10+2$ dimensional target space--time 
for superstring. 
Both string models have two time coordinates. 
In order to construct 
the covariant Lagrangians of the {\UvUa} models 
the generalized Chern-Simons term plays an important role. 
The supersymmetric generalized Chern-Simons action is 
also proposed. 
The Green-Schwarz type of {\UvUa} superstring model has 
another fermionic local symmetry as well as $\kappa$-symmetry.
The supersymmetry of target space--time is different from 
the standard one.

\vspace{12pt}


\section{Introduction}\label{sec:introduction}

The motivation of this paper is to understand the reason why 
there is only one time coordinate in real world.
If there are more than two time coordinates, what happens?\ \ 
To consider more than two time coordinates might be 
a clue to understand the origin of time.
In order to understand this problem 
we in this paper study 
the symmetric structure of several {\UvUa} models 
each of which has an extra time coordinate.

There are several models 
which has more than two time coordinates. 
For example, 
F-theory \cite{Ftheory}, 
two-time physics \cite{Bars1,Bars2}, 
and so on. 
F-theory is constructed as a field theory of (2,2)-brane 
in $10+2$ dimensional space--time, 
i.e.\ 10 space coordinates and 2 time coordinates. 
The two-time physics proposed by I.\ Bars et.\ al.\ is 
constructed by using the field theory of multi particles. 
Several years ago, the author also proposed string models 
which have two time coordinates in ref.~\cite{U1U1string}.
These models have {\UvUa} gauge symmetry in two dimensions. 
The target space--time of {\UvUa} bosonic string is 
$26+2$ dimensions, 
while 
that of {\UvUa} superstring is $10+2$ dimensions. 
The manifest covariant expression of these models 
was not found in those days. 
In this paper we will give the covariant Lagrangians 
and will study the local gauge symmetries. 

The {\UvUa} models are field theories 
defined in two-dimensional space--time.
In section 2 we study the {\UvUa} bosonic models 
without and with two-dimensional gravity.
In subsection 2.1 
we introduce the {\UvUa} bosonic model 
without gravity \cite{U1U1model}
and extend this model to the model which has manifestly 
ISO($D-1$,1) Poincar\'e symmetry. 
In order to obtain the ISO($D-1$,1) symmetry, 
i.e.\ 
in order to obtain the covariant expression,
the generalized Chern-Simons term plays an important role. 
In subsection 2.2 
the property of the generalized Chern-Simons term 
introduced in the previous subsection is discussed. 
In subsection 2.3 we couple the {\UvUa} model 
to two-dimensional gravity, i.e.\ 
we obtain a kind of string theory 
which has {\UvUa} symmetry in two dimensions and 
ISO(26,2) Poincar\'e symmetry at the target space--time. 
Though this model is already proposed 
in ref.~\cite{U1U1string}, 
the covariant expression given in this paper is new. 
In section 3 we study the supersymmetric version of 
the {\UvUa} models without and with two-dimensional supergravity. 
We covariantize the Lagrangian of 
the Neveu-Schwarz-Ramond type of {\UvUa} superstring model 
proposed in ref.~\cite{U1U1string}.
This superstring model has 
ISO(10,2) Poincar\'e symmetry in the target space--time. 
The supersymmetric version of 
the generalized Chern-Simons Lagrangian, 
which is necessary for the covariantization of {\UvUa} models, 
is proposed here. 
Section 4 is devoted to the Green-Schwarz type 
of {\UvUa} superstring model. 
We will show that this model has 
another fermionic local symmetry besides $\kappa$-symmetry. 
Section 5 is the discussions and the conclusions. 


\section{{\UvUa} Bosonic Theories}
\label{sec:Bosonic_Theories}

\subsection{{\UvUa} bosonic model without gravity}
\label{sec:bosonic_model}

In this subsection we study the {\UvUa} model \cite{U1U1model} 
which is described in two dimensions 
and has no gravitational field. 
The space--time coordinates are $x^m$ ($m = 0$, $1$) and 
the signature of metric $\eta_{mn}$ is $(-,+)$. 
The fields of the {\UvUa} model 
consists of an Abelian gauge field $A_m(x)$ 
and several matter fields. 
The theory has 
both a vector U(1) local symmetry \lq\lq ${\rm U}(1)_{\rm V}$'' 
and an axial vector U(1) local symmetry 
\lq\lq ${\rm U}(1)_{\rm A}$'' 
in two dimensions, 
i.e.\ the theory is invariant 
under the local gauge transformation, 
$\delta A_m(x) = \prt_m v(x) + \ep_m{}^n \prt_n \tv(x)$. 

In refs.\ \cite{U1U1model} the {\UvUa} model is proposed. 
The Lagrangian is 
\beq{LagrangianLCfermionic}
\cL \ = \  
- i \, \sum_a \psi_a \sg^m \prt_m \psi_a 
- \, \sum_a e_a A_m \, j^{a m} 
+ \pi \sum_{a,b} g_{a b} \, j^{a m} j_m^b
\, ,
\eeq
and the current is 
\beq{CurrentLCfermionic}
j^{a m}  \ = \  \psi_a \sg^m \psi_a
\, ,
\eeq
where the indices $a$ and $b$ run 
$1$, $2$, $\ldots$, $D$ ($D \ge 2$). 
$\psi_a$ are spinor fermionic matter fields. 
The charges $e_a$ and the coupling constants $g_{ab}$ 
satisfy the following two conditions,
\bea
&&
\hbox{{\bf a)}~~~} 
\sum_{a,b} G^{ab} \, e_a e_b  \ = \  0 \, ,
\label{LightlikeCondition}\\
&&
\hbox{{\bf b)}~~~} 
\hbox{one of the eigenvalues of $G_{ab}$ is negative, 
      others are positive}
\label{NoGhostCondition}
\eea
where
\beq{BackgroundMetric}
G_{ab}  \ = \  \delta_{ab} + 2 g_{ab} \, , 
~~~~~~~~
G^{ab}  \ = \  (G_{ab})^{-1} \, .
\eeq
Both coupling constants $g_{ab}$ and $G_{ab}$ are 
real symmetric matrices. 
The Lagrangian \rf{LagrangianLCfermionic} is invariant 
under the following {\it local}\/ transformations, 
\bea
\delta A_m     &=&  \prt_m v + \ep_m{}^n \prt_n \tv \, ,
\label{U1U1transfA}\\
\delta \psi_a  &=&  i e_a (v - \bar{\sg} \tv) \psi_a \, .
\label{U1U1transfpsi}
\eea
The parameters $v(x)$ and $\tv(x)$ describe 
${\rm U}(1)_{\rm V}$ and ${\rm U}(1)_{\rm A}$ 
gauge transformations, respectively. 
In the model defined by 
the Lagrangian \rf{LagrangianLCfermionic}, 
${\rm U}(1)_{\rm V}$ gauge symmetry is manifest, 
on the other hand, ${\rm U}(1)_{\rm A}$ gauge symmetry 
is non-trivial because 
the axial-vector anomaly is cancelled non-perturbatively. 
The non-perturbative cancellation of the axial-vector anomaly 
requires the condition \rf{LightlikeCondition}. 
This condition leads us to the fact that 
the matrix $G_{ab}$ is not positive definite 
nor negative definite. 
On the other hand, 
the unitarity, i.e.\ the absence of negative norm states, 
requires the condition that 
the number of negative eigenvalues of $G_{ab}$ 
is equal to zero or one.
Together with condition \rf{LightlikeCondition}, 
the condition \rf{NoGhostCondition} is obtained. 

The bosonization leads us to the Lagrangian 
\beq{LagrangianLCbosonic}
\cL \ = \  
- \, \half \sum_{a,b} G_{ab} \prt^m \xi^a \prt_m \xi^b 
- \, \sum_a e_a A_m \, j^{a m} 
\, ,
\eeq
and the current
\beq{CurrentLCbosonic}
j^{a m}  \ = \  \frac{1}{\sqrt{\pi}} \, \ep^{mn} \prt_n \xi^a
\, ,
\eeq
where $\xi^a$ are scalar bosonic matter fields. 
The Lagrangian \rf{LagrangianLCbosonic} is invariant 
under the local transformations \rf{U1U1transfA} and 
\beq{U1U1transfxi}
\delta \xi^a  \ = \  
\frac{1}{\sqrt{\pi}} \sum_b G^{ab} e_b \, \tv
\, .
\eeq
Both ${\rm U}(1)_{\rm V}$ and ${\rm U}(1)_{\rm A}$ symmetries 
are manifest in the bosonized expression. 

In the Lagrangian \rf{LagrangianLCbosonic}, 
$G_{ab}$ can be considered to be a background metric. 
So, it is decomposed by using 
the $D$-dimensional vielbein $\zeta_a^{I}$ 
which satisfies 
\beq{BackgroundMetricVielbein}
G_{ab}  \ = \  
\sum_{I,J} \eta_{IJ} \, \zeta_a^{I} \zeta_b^{J}  \, ,
~~~~~~~~
\sum_{a,b} G^{ab} \zeta_a^{I} \zeta_b^{J} = \eta^{IJ} \, ,
\eeq
where $\eta_{IJ}$ is a symmetric matrix defined by 
\beq{FlatmetricDefinition}
\eta_{IJ}  \ = \  \eta^{IJ}  \ = \ 
    \left\{\begin{array}{cl}
           1   &  \hbox{($I=J=i$, \ $i=1$, $2$, \ldots, $D-2$)} \\
          -1   &  \hbox{($I=J=\widehat{0}$)} \\
           1   &  \hbox{($I=J=\widehat{1}$)} \\
           0   &  \hbox{(otherwise)}
    \end{array}\right.
\, .
\eeq
The index $i$ runs $1$, $2$, $\ldots$, $D-2$, 
while 
the index $I$ runs $1$, $2$, $\ldots$, $D-2$, 
and $\widehat{0}$, $\widehat{1}$. 
In our notation, the indices $I$ and $J$ 
will always run over the above all values. 
The negative metric $\eta_{\widehat{0}\widehat{0}}$ 
comes from the property \rf{NoGhostCondition}.
The light-cone-like indices $\widehat{+}$ and $\widehat{-}$ 
are introduced by 
\beq{plusminussign}
X^{\widehat{\pm}}  \ = \  
\frac{1}{\sqrt{2}} ( X^{\widehat{0}} \pm X^{\widehat{1}} )
\, .
\eeq
Then, one has, for example, 
$\eta_{\widehat{+}\widehat{-}} = \eta_{\widehat{-}\widehat{+}} =
 \eta^{\widehat{+}\widehat{-}} = \eta^{\widehat{-}\widehat{+}} = -1$ 
and 
$\eta_{\widehat{+}\widehat{+}} = \eta_{\widehat{-}\widehat{-}} =
 \eta^{\widehat{+}\widehat{+}} = \eta^{\widehat{-}\widehat{-}} = 0$. 
Without loss of generality, one can choose $\zeta_a^{\widehat{-}}$ as 
\beq{e_zeta}
\zeta_a^{\widehat{-}}  =  e_a  \, ,
\eeq
owing to the condition \rf{LightlikeCondition}.
Then, the Lagrangian \rf{LagrangianLCbosonic} is rewritten as 
\beq{LagrangianLC}
\cL \ = \  
- \, \half \, \prt^m \xi^I \prt_m \xi_I 
+ \frac{1}{\sqrt{\pi}} \tA^m \prt_m \xi^{\widehat{-}} 
\, ,
\eeq
where 
\beq{}
\xi^I  \ = \  \sum_a \zeta_a^I \, \xi^a
\eeq
and $\tA^m = \ep^{mn} A_n$. 
The indices $I$ and $J$ are lowered and raised 
by the background metric $\eta_{IJ}$ and $\eta^{IJ}$, respectively. 

Using the light-cone-like indices, 
the local transformation \rf{U1U1transfxi} becomes 
\beq{U1U1transfxiLC}
\delta \xi^{\widehat{+}}  \ = \  - \, \frac{1}{\sqrt{\pi}} \, \tv 
\, , ~~~~~~~~
\delta \xi^I  \ = \  0  ~~~\hbox{($I \neq \widehat{+}$)}
\, .
\eeq
This transformation makes us possible to take 
the light-cone-like gauge fixing condition, 
\beq{LightConeBosonic}
\xi^{\widehat{+}}  \ = \  0 \, .
\eeq
The path-integration of $\tA^m$ gives 
\beq{LightconeBosonic}
\prt_m \xi^{\widehat{-}} = 0
\, . 
\eeq
The gauge field $\tA^m$ plays a role of a Lagrange multiplier. 


The first term of the Lagrangian \rf{LagrangianLC} has 
Poincar\'e ISO($D-1$,1) global symmetry. 
So it is quite natural that one expects 
the existence of the Lagrangian 
which has ISO($D-1$,1) global symmetry. 
Introducing the auxiliary field $\auxA^I(x)$ 
and new gauge fields 
$\tB^{mI}(x) = \ep^{mn} B^I_n(x)$ and 
$\tC(x) = \half \ep^{mn} C_{mn}(x)$,
we have succeeded in extending the Lagrangian \rf{LagrangianLC} 
to the Lagrangian which has ISO($D-1$,1) global symmetry, 
\beq{LagrangianBosonic}
\cL \ = \ 
- \, \half \prt^m \xi^I \prt_m \xi_I 
+ \tA^m \auxA_I \prt_m \xi^I
+ \tB^{mI} \prt_m \auxA_I
- \half \tC \auxA^I \auxA_I  \, .
\eeq
The Lagrangian \rf{LagrangianBosonic} is invariant 
under the following local gauge transformations:
\bea
\delta \xi^I &=&   \tv \auxA^I
\, , \nonumber\\
\delta \tA_m &=&   \prt_m \tv + \ep_m{}^n \prt_n v
\, , \nonumber\\
\delta \auxA^I &=& 0
\, , \label{LocalGaugeTransfBosonic}\\
\delta \tB_m^I &=& 
                   \ep_m{}^n \prt_n u^I + \tv \prt_m \xi^I 
                   - v \ep_m{}^n \prt_n \xi^I - \tw_m \auxA^I
\, , \nonumber\\
\delta \tC &=&     \prt_m \tw^m + \prt_m \tv \tA^m - \tv \prt_m \tA^m
\, , \nonumber
\eea
where $u^I(x)$ and $\tw_m(x) = \ep^{mn} w_n(x)$ 
are new local gauge parameters. 
It should be noted that 
the gauge transformation \rf{LocalGaugeTransfBosonic} 
has on-shell reducibility, 
i.e.\ 
the transformation \rf{LocalGaugeTransfBosonic} 
has on-shell invariance under the following local gauge transformation, 
\bea
\delta'  v    &=& 0
\, , \nonumber\\
\delta' \tv   &=& 0
\, , \label{LocalGaugeTransfBosonic2}\\
\delta' u^I   &=& w' \auxA^I
\, , \nonumber\\
\delta' \tw_m &=& \ep_m{}^n \prt_n w' 
\, . \nonumber
\eea
The Lagrangian \rf{LagrangianBosonic} is also invariant 
under the global transformations of ISO($D-1$,1) and scale symmetries 
as follows: 
\bea
\delta \xi^I  &=&  \omega^I{}_J \xi^J  +  a^I
\, , \nonumber\\
\delta \tA_m  &=&  r \tA_m + \sum_{i=1}^{2g} \alpha_i h^{(i)}_m
\, , \nonumber\\
\delta \auxA^I &=& - \, r \auxA^I + \omega^I{}_J \auxA^J
\, , \label{GlobalGaugeTransfBosonic}\\
\delta \tB_m^I &=& 
  r \tB_m^I + \omega^I{}_J \tB_m^J  
  + \sum_{i=1}^{2g} ( \beta_i^I + \alpha_i \xi^I) h^{(i)}_m
\, , \nonumber\\
\delta \tC &=& 2r \tC 
\, , \nonumber
\eea
where $\omega_{IJ} = -\omega_{JI}$, 
$a^I$, $r$, $\alpha_i$ and $\beta_i^I$ are global parameters, 
and $h^{(i)}_m(x)$ 
are harmonic functions which satisfy 
$\prt^m h^{(i)}_m = \ep^{mn} \prt_m h^{(i)}_n = 0$ 
($i=1,2,\ldots,2g$; $g=$ genus of 2D space--time). 
The Lagrangian \rf{LagrangianBosonic} is also invariant 
under the following two discrete transformations:
\bea
&&\hbox{i)}  \hspace{12pt}  \hbox{
           $\xi^I \rightarrow -\xi^I$, 
           $\auxA^I \rightarrow -\auxA^I$,
           $\tB_m^I \rightarrow -\tB_m^I$,
           otherwise unchanged,}    
\label{DiscreteTransfBosonic1}\\
&&\hbox{ii)}  \hspace{8pt}  \hbox{
           $\tA_m \rightarrow -\tA_m$, 
           $\auxA^I \rightarrow -\auxA^I$,
           $\tB_m^I \rightarrow -\tB_m^I$,
           otherwise unchanged.} 
\label{DiscreteTransfBosonic2}
\eea
Both discrete transformations are 
a kind of parity transformations of background space--time.

The equations of motion of the Lagrangian \rf{LagrangianBosonic} 
are 
\bea
&& \prt^m \prt_m \xi^I - \prt_m ( \tA^m \auxA^I )  \ = \  0  
\, , \nonumber\\
&& \auxA_I \prt_m \xi^I  \ = \  0  
\, , \nonumber\\
&& \tA^m \prt_m \xi^I - \prt_m \tB^{mI} - \tC \auxA^I  \ = \  0  
\, , \label{EOMBosonic}\\
&& \prt_m \auxA^I  \ = \  0  
\, , \nonumber\\
&& \auxA^I \auxA_I  \ = \  0  
\, . \nonumber
\eea
The solution of 
$\prt_m \auxA^I = 0$ and $\auxA^I \auxA_I = 0$ 
in \rf{EOMBosonic} is 
\beq{auxiliaryfieldSolution}
\auxA^I  \ = \ 
    \left\{\begin{array}{cl}
           0      &  \hbox{($I=i$)}     \\
      - \, \frac{1}{\sqrt{\pi}}     &  \hbox{($I=\widehat{+}$)} \\
           0      &  \hbox{($I=\widehat{-}$)} 
    \end{array}\right.  \, ,
\eeq
where we have used the global transformations 
\rf{GlobalGaugeTransfBosonic}. 
Though $\auxA^I = 0$ is another solution, 
one needs a concept of fine tuning to realize this solution. 
In the case of the solution \rf{auxiliaryfieldSolution}, 
the gauge transformation \rf{LocalGaugeTransfBosonic} 
makes us possible to choose 
the light-cone-like gauge fixing condition \rf{LightConeBosonic}, 
and then, 
the Lagrangian \rf{LagrangianBosonic} becomes \rf{LagrangianLC}. 
Thus, the equivalence of the Lagrangians 
\rf{LagrangianLC} and \rf{LagrangianBosonic} has been proved 
at the classical level.

\subsection{Generalized Chern-Simons action}
\label{sec:GCS_bosonic_action}

In this subsection we will show that 
a part of the Lagrangian \rf{LagrangianBosonic}, 
\beq{LagrangianGCSoriginal}
\cL_{\rm GCS} \ = \  
\tB^{mI} \prt_m \auxA_I
- \half \tC \auxA^I \auxA_I 
\, ,
\eeq
which is introduced 
for covariantizing the Lagrangian \rf{LagrangianLC},
is the generalized Chern-Simons Lagrangian \cite{kawata}. 

The generalized Chern-Simons Lagrangian in two dimensions is defined by 
\cite{kawata}
\beq{LagrangianGCS}
\cL_{\rm GCS} \ = \  
\Htr \Bigl\{ \auxA ( \dd B + B^2 ) + \auxA^2 C \Bigr\}
\, ,
\eeq
where 
\beq{GCSrep}
\auxA   \ = \  \half \Sigma_{\hat \a} \auxA^{\hat \a}     \, ,
~~~~~~
B       \ = \  \half \dd x^m T_{\hat a} B_m^{\hat a}      \, ,
~~~~~~
C       \ = \  \quarter \dd x^m \!\!\wedge\! \dd x^n \, 
               \Sigma'_{\hat \a'} C_{mn}^{\hat \a'}       \, ,
\eeq
and $\dd = \dd x^m \prt_m$. 
$\Htr$ is a trace defined below. 
The Lagrangian \rf{LagrangianGCS} has the gauge invariance 
under the local gauge transformation:
\bea
\delta \auxA &=&  [ \auxA , u ]   \, ,
\nonumber\\
\delta B     &=&  \dd u + [ B , u ] - \{ \auxA , w \}  \, ,
\label{GCStransf}\\
\delta C     &=&  \dd w + \{ B , w \} + [ C , u ] - [ \auxA , q ]  \, ,
\nonumber
\eea
where
\beq{GCSrep2}
u   \ = \  \half T_{\hat a} u^{\hat a}        \, ,
~~~~~~
w   \ = \  \half \dd x^m \Sigma'_{\hat \a'} w_m^{\hat \a'}  \, ,
~~~~~~
q       \ = \  \quarter \dd x^m \!\!\wedge\! \dd x^n \, 
                T'_{\hat a'}q_{mn}^{\hat a'}      \, .
\eeq
Here, $T$, $T'$, $\Sigma$ and $\Sigma'$ satisfy 
a graded Lie algebra, 
\bea
&&\hspace{-3pt}
\{ \Sigma_{\hat \a} , \Sigma'_{\hat \b'} \}  =  
f_{\hat \a \hat \b'}{}^{\hat c} T_{\hat c} 
\, , \label{gradedLie}\\
&&
[ \Sigma_{\hat \a} , T_{\hat b} ]  =  
f_{\hat \a \hat b}{}^{\hat c} \Sigma_{\hat c}  \, ,
~~~~~~
[ \Sigma'_{\hat \a'} , T_{\hat b} ]  =  
f_{\hat \a' \hat b}{}^{\hat \g'} \Sigma'_{\hat \g'} 
~~~~~~
[ \Sigma_{\hat \a} , T'_{\hat b'} ]  =  
f_{\hat \a \hat b'}{}^{\hat \g'} \Sigma'_{\hat \g'} 
\, , \nonumber\\
&&
[ T_{\hat a} , T_{\hat b} ]  =  
f_{\hat a \hat b}{}^{\hat c} T_{\hat c}  \, ,
\hspace{29pt}
[ T_{\hat a} , T'_{\hat b'} ]  =  
f_{\hat a \hat b'}{}^{\hat c'} T'_{\hat c'}  
\, . \nonumber
\eea
Note that 
$\Sigma_{\hat \a}$ and $\Sigma'_{\hat \a'}$ 
are not Grassmannian. 
In $\Htr$ one can move any 
$T$, $T'$, $\Sigma$ or $\Sigma'$ 
cyclically, 
i.e.\ 
$\Htr( T \ldots ) = \Htr ( \ldots T )$, 
$\Htr( \Sigma \ldots ) = \Htr ( \ldots \Sigma )$, 
and so on.

In order to obtain \rf{LagrangianGCSoriginal} from \rf{LagrangianGCS}, 
one chooses the representation, 
\bea
&&
\Sigma_{\hat  \a}  =  \Gamma_I , ~ \sg_+ [ \Gamma_I , \Gamma_J ]  \, ,
~~~~~~
T_{\hat a}        =  \sg_+ \Gamma_I  \, ,
\label{GCSrepresentation}\\
&&
\Sigma'_{\hat \a'}  =  \sg_+  \, ,
\hspace{79pt}
T'_{\hat a'}  =  \sg_+  \, ,
\nonumber
\eea
i.e.\ 
\bea
\auxA &=& \half \Gamma_I \auxA^I + 
          \quarter \sg_+ [ \Gamma_I , \Gamma_J ] 
          \auxA^{\prime IJ}
\, , \nonumber\\
B &=& \half \sg_+ \Gamma_I B^I_m \dd x^m
\, , \label{GCSrep3}\\
C &=& \quarter \sg_+ C_{mn} \dd x^m \!\!\wedge\! \dd x^n
\, , \nonumber
\eea
and
\bea
u &=& \half \sg_+ \Gamma_I u^I 
\, ,
\nonumber\\
w &=& \half \sg_+ w_m \dd x^m
\, ,
\label{GCSrep4}\\
q &=& \quarter \sg_+ q_{mn} \dd x^m \!\!\wedge\! \dd x^n
\, ,
\nonumber
\eea
where 
$B^I_m = \ep_{mn} \tB^{nI}$, $C_{mn} = - \ep_{mn} \tC$, 
and $w_m = \ep_{mn} \tw^n$. 
$\sg_+$ and $\Gamma_I$ are the gamma matrices 
of SO(1,1) and SO($D-1$,1) respectively, 
and satisfy 
$\{ \sg_m , \sg_n \} = 2 \eta_{mn}$ 
and 
$\{ \Gamma_I , \Gamma_J \} = 2 \eta_{IJ}$ 
respectively. 
$\sg_m$ is a $2 \times 2$ matrix 
while
$\Gamma_I$ is a $2^{D/2} \times 2^{D/2}$ matrix. 
The trace $\Htr$ is defined by 
$\Htr( \cdots ) = \frac{-1}{2^{(D-2)/2}} \tr (\sg_- \cdots )$, 
where $\tr$ is a usual trace which takes 
both $\sg$-matrix and $\Gamma$-matrix into account, 
for example, $\tr 1 = 2 \cdot 2^{D/2}$. 
The concrete expression for $\sg$-matrices is 
\beq{gamma_matrix}
\sg_+ \ = \ 
\left( \begin{array}{cc}
           0 & 0 \\
    \sqrt{2} & 0
       \end{array}
\right) \, ,  \qquad
\sg_- \ = \ 
\left( \begin{array}{cc}
           0 & -\sqrt{2} \\
           0 & 0
       \end{array}
\right) \, .
\eeq
One finds, for example, 
$\Htr \sg_+ = 4$, 
$\Htr ( \sg_+ \Gamma_I \Gamma_J ) = 4 \eta_{IJ}$.

Since $B^2=0$, $[B,u]=0$, $\{B,w\}=0$, $[C,u]=0$ and $[\auxA,q]=0$
in the case of representation \rf{GCSrep3}, 
the Lagrangian \rf{LagrangianGCS} becomes a simple form, 
\beq{LagrangianGCS3}
\cL_{\rm GCS} \ = \  
\Htr \bigl( \auxA \hspace{1pt} \dd B + \auxA^2 C \bigr)
\, ,
\eeq
and the gauge transformation \rf{GCStransf} becomes 
\bea
\delta \auxA &=& [ \auxA , u ] 
\, ,
\nonumber\\
\delta B &=& \dd u - \{ \auxA , w \}
\, ,
\label{GCStransf2}\\
\delta C &=& \dd w 
\, .
\nonumber
\eea
The field $\auxA^{\prime IJ}$ is decoupled 
because it does not appear in the Lagrangian. 
Therefore, one can set $\auxA^{\prime IJ} = 0$ without loss of generality, 
namely, 
$\auxA = \half \Gamma_I \auxA^I$ and 
the gauge transformation of $\auxA$ is 
$\delta \auxA = 0$. 
The gauge parameter $q_{mn}$ is also decoupled.

\subsection{{\UvUa} bosonic string}\label{sec:bosonic_string}

In this subsection we consider 
the {\UvUa} bosonic model coupled to 2D gravity, i.e.\ 
the {\UvUa} bosonic string model. 

The {\UvUa} bosonic string model is described 
by the Lagrangian \rf{LagrangianBosonic} 
under the curved 2D space--time with metric $g_{mn}$, i.e.\ 
\beq{LagrangianBosonicGravity}
\cL \ = \  \sqrt{-g} \, \Big(
- \half g^{mn} \prt_m \xi^I \prt_n \xi_I 
+ \tA^m \auxA_I \prt_m \xi^I
+ \tB^{mI} \prt_m \auxA_I
- \half \tC \auxA^I \auxA_I  \Big)
\, ,
\eeq
where $g(x) = \det g_{mn}(x)$. 
In this model 
the unitarity allows us two negative eigenvalues of $G_{ab}$, 
i.e.\ the condition 
\bea
&&
\hbox{{\bf b${}'$)}~~~} 
\hbox{two of the eigenvalues of $G_{ab}$ is negative, 
      others are positive}
\label{NoGhostCondition2}
\eea
is required instead of the condition \rf{NoGhostCondition} 
because the general coordinate transformations 
as well as the ${\rm U}(1)_{\rm A}$ gauge transfomations 
remove one negative norm state, respectively. 
Therefore, 
the index $I$ runs 
$0$, $1$, $2$, \ldots, $D-3$, $\widehat{0}$, $\widehat{1}$, 
where one has to use the background metric $\eta_{IJ}$ 
\beq{FlatmetricDefinitionGravity}
\eta_{IJ}  \ = \  \eta^{IJ}  \ = \ 
    \left\{\begin{array}{cl}
          -1   &  \hbox{($I=J=0$)}           \\
           1   &  \hbox{($I=J=i$, \ $i=1$, $2$, \ldots, $D-3$)}  \\
          -1   &  \hbox{($I=J=\widehat{0}$)} \\
           1   &  \hbox{($I=J=\widehat{1}$)} \\
           0   &  \hbox{(otherwise)}
    \end{array}\right.
\eeq
instead of \rf{FlatmetricDefinition}. 
There are two negative metric components 
$\eta_{00}$ and $\eta_{\widehat{0}\widehat{0}}$, 
while there is one negative metric component 
$\eta_{\widehat{0}\widehat{0}}$ 
in the {\UvUa} bosonic model without gravity. 

This Lagrangian is invariant 
under the local {\UvUa} transformations, 
the general coordinate transformations and the Weyl scaling as follows:
\bea
\delta \xi^I &=&   \tv \auxA^I
\, , \nonumber\\
\delta \tA_m &=&   \prt_m \tv + \EP_m{}^n \prt_n v 
\, , \nonumber\\
\delta \auxA^I &=& 0
\, , \label{LocalGaugeTransfBosonicGravity}\\
\delta \tB_m^I &=& 
                   \EP_m{}^n \prt_n u^I + \tv \prt_m \xi^I 
                   - v \, \EP_m{}^n \prt_n \xi^I - \tw_m \auxA^I
\, , \nonumber\\
\delta \tC &=&     \nabla\!_m \tw^m + \prt_m \tv \tA^m - \tv \nabla\!_m \tA^m
\, , \nonumber\\
\delta g_{mn} &=&  0
\, , \nonumber
\eea
and
\bea
\delta \xi^I &=&   k^n \prt_n \xi^I
\, , \nonumber\\
\delta \tA_m &=&   k^n \prt_n \tA_m + \prt_m k^n \tA_n
\, , \nonumber\\
\delta \auxA^I &=& k^n \prt_n \auxA^I
\, , \label{GeneralCoordinateTransfBosonicGravity}\\
\delta \tB_m^I &=& k^n \prt_n \tB_m^I + \prt_m k^n \tB_n^I
\, , \nonumber\\
\delta \tC &=&     k^n \prt_n \tC + 2 s \tC
\, , \nonumber\\
\delta g_{mn} &=&  k^l \prt_l g_{mn} + \prt_m k^l g_{ln} + \prt_n k^l g_{ml}
                   - 2 s g_{mn}
\, , \nonumber
\eea
where $k^n(x)$ and $s(x)$ are local parameters for 
the general coordinate transformation and the Weyl transformation. 
$\EP^{mn}(x) = \ep^{mn} / \sqrt{-g(x)}$ 
is the anti-symmetric tensor on 2D curved space--time. 
It should be noted that 
the gauge transformation \rf{LocalGaugeTransfBosonicGravity} 
has on-shell reducibility, 
as the same as the gauge transformation \rf{LocalGaugeTransfBosonic}.
The transformation \rf{LocalGaugeTransfBosonicGravity} 
has on-shell invariance under the local gauge transformation, 
\bea
\delta'  v    &=& 0
\, , \nonumber\\
\delta' \tv   &=& 0
\, , \label{LocalGaugeTransfBosonicGravity2}\\
\delta' u^I   &=& w' \auxA^I
\, , \nonumber\\
\delta' \tw_m &=& \EP_m{}^n \prt_n w' 
\, , \nonumber
\eea
which is the two-dimensional covariant form of 
\rf{LocalGaugeTransfBosonic}. 

The Lagrangian \rf{LagrangianBosonicGravity} is also invariant 
under the global transformations \rf{GlobalGaugeTransfBosonic} 
and the discrete transformations 
\rf{DiscreteTransfBosonic1} and \rf{DiscreteTransfBosonic2}, 
where $h^{(i)}_m(x)$ 
are harmonic functions on curved 2D space--time 
which satisfy 
$\nabla^m h^{(i)}_m = \EP^{mn} \nabla\!_m h^{(i)}_n = 0$ 
($i=1,2,\ldots,2g$). 
The Poincar\'e symmetry is ISO($D-2$,2). 

By introducing new fields 
$\hat{A}^m$, $\hat{B}^{mI}$, $\hat{C}$, $\hat{g}^{mn}$, $p$, 
and a new parameter $\hat{w}^m$, 
\bea
&&
\hat{A}^m     \ = \  \sqrt{-g} \, \tA^m
\, , ~~~~~~
\hat{B}^{mI}  \ = \  \sqrt{-g} \, \tB^{mI}
\, , ~~~~~~
\hat{C}       \ = \  \sqrt{-g} \, \tC
\, , \\
&&
\hat{g}^{mn}  \ = \  \sqrt{-g} \, g^{mn}
\, , ~~~~~
p
\, , \hspace{111pt}
\hat{w}^m     \ = \  \sqrt{-g} \, \tw^m
\nonumber
\eea
the Lagrangian \rf{LagrangianBosonicGravity} is rewritten as 
\bea
\cL  &=&
- \, \half \hat{g}^{mn} \prt_m \xi^I \prt_n \xi_I 
+ \hat{A}^m \auxA_I \prt_m \xi^I
+ \hat{B}^{mI} \prt_m \auxA_I
- \half \hat{C} \auxA^I \auxA_I
\label{LagrangianBosonicGravityModified}\\
&&+ \, p \, ( \det\!\hat{g}^{mn} + 1 )
\, ,
\nonumber
\eea
and the gauge transformations 
\rf{LocalGaugeTransfBosonicGravity} and 
\rf{GeneralCoordinateTransfBosonicGravity} are rewritten as 
\bea
\delta \xi^I &=&   k^n \prt_n \xi^I  +  \tv \auxA^I
\, , \nonumber\\
\delta \hat{A}^m &=&  \prt_n ( k^n \hat{A}^m ) - \prt_n k^m \hat{A}^n + 
                      \hat{g}^{mn} \prt_n \tv + \ep^{mn} \prt_n v 
\, , \nonumber\\
\delta \auxA^I  &=&  k^n \prt_n \auxA^I
\, , \label{LocalGaugeTransfBosonicGravityModified}\\
\delta \hat{B}^{mI} &=& 
                \prt_n ( k^n \hat{B}^{mI} ) - \prt_n k^m \hat{B}^{nI}
              + \ep^{mn} \prt_n u^I 
\nonumber\\
           && + \, \tv \hat{g}^{mn} \prt_n \xi^I 
              - v \, \ep^{mn} \prt_n \xi^I - \hat{w}^m \auxA^I
\, , \nonumber\\
\delta \hat{C} &=& \prt_n ( k^n \hat{C} ) + 
      \prt_m \hat{w}^m + \prt_m \tv \hat{A}^m - \tv \prt_m \hat{A}^m
\, , \nonumber\\
\delta \hat{g}^{mn} &=&  \prt_l ( k^l \hat{g}^{mn} ) 
      - \prt_l k^m \hat{g}^{ln} - \prt_l k^n \hat{g}^{ml}
\, , \nonumber\\
\delta p  &=&  \prt_n ( k^n p ) 
\, , \nonumber
\eea
where $p(x)$ is a Lagrange multiplier field. 
The third and the fourth terms 
in \rf{LagrangianBosonicGravityModified} 
is the generalized Chern-Simons term 
$\cL_{\rm GCS}$ \rf{LagrangianGCSoriginal}. 
Both 
the Lagrangian \rf{LagrangianBosonicGravityModified} 
and 
the gauge transformations \rf{LocalGaugeTransfBosonicGravityModified}
are polynomials of fields and parameters. 

At the quantum level, 
the absence of conformal anomaly requires $D=28$. 
We will give the reason in the following. 
The conformal charge for a spin $j$ particle is 
$6j^2 - 6j + 1$. 
The gauge field give a negative sign to the conformal charge 
because the FP ghost field has opposite statistics. 
The {\UvUa} bosonic string model consists of 
one graviton ($j=2$), one photon ($j=1$), 
and $D$ scalar particles ($j=0$). 
Therefore, 
the total conformal charge of {\UvUa} bosonic string model is 
\bea
w^{(N=0)} 
&=&  2 \!\times\! (-13) + 2 \!\times\! (-1) + D \!\times\! 1 
\label{ConformalWeightBosonic}\\
&=&  D - 28 
\, , \nonumber
\eea
where $2$ comes from the number of components. 
The fields $\auxA^I$, $\tB_m^I$, and $\tC$ 
do not contribute to the conformal charge 
because these fields come from the generalized Chern-Simons action 
which is considered to be topological. 
The cancellation of conformal anomaly, i.e.\ 
the existence of the Weyl symmetry at the quantum level 
requires $w^{(N=0)} = 0$. 
Thus, we obtain $D=28$ from \rf{ConformalWeightBosonic}. 
The detail calculation about the quantization 
will be given in the paper \cite{tsukiwata}. 
The discussion about the spectrum will be also given there.

\section{{\UvUa} Supersymmetric Theories}
\label{sec:Supersymmetric_Theories}

In this section we extend the previous models to 
the models with $N=1$ supersymmetry. 
We use the (1,1) type superspace with coordinates 
$z^M = (x^m, \th^\m)$ ($m = 0$, $1$; $\m = 1$, $2$). 
$\theta^\mu$ are fermionic spinor coordinates. 

\subsection{{\UvUa} supersymmetric model without gravity}
\label{sec:supersymmetric_model}

In the case of supersymmetric version of 
the {\UvUa} model without gravity, 
we introduce superfields 
$\Xi^I(z)$, $\tAA^\a(z) = (\bar{\sg}\Psi(z))^\a$, 
$\AuxA^I(z)$, $\tBB^{\a I}(z) = (\bar{\sg}\Pi^I(z))^\a$ 
and $\tCC(z) = - \half \bar{\sg}^{\a\b} \L_{\a\b}(z)$, 
instead of 
$\xi^I(x)$, $\tA^m(x)$, $\auxA^I(x)$, $\tB^{m I}(x)$ 
and $\tC(x)$, 
respectively. 
$\Xi^I$, $\AuxA^I$ and $\tCC$ are 
scalar bosonic superfields, on the other hand, 
$\tAA^\a$ and $\tBB^{\a I}$ are 
spinor fermionic superfields. 
In ref.~\cite{U1U1string} 
the supersymmetric version of 
\rf{LagrangianLCbosonic} is proposed. 
In the previous section we have succeeded in covariantizing 
the Lagrangian of the {\UvUa} bosonic model. 
So, it is now easy to obtain the covariant Lagrangian of 
the {\UvUa} supersymmetric model. 
The covariant Lagrangian for the {\UvUa} supersymmetric model 
without gravity is 
\beq{LagrangianSuper}
\cL \ = \  
- \, \half \bDD^\a \Xi^I \bDD_\a \Xi_I 
+ \tAA^\a \AuxA_I \bDD_\alpha \Xi^I 
+ \tBB^{\a I} \bDD_\a \AuxA_I
- \half \tCC \AuxA^I \AuxA_I   
\, .
\eeq
The background metric $\eta_{IJ}$ is \rf{FlatmetricDefinition}. 
This Lagrangian is invariant 
under the following local and global transformations:
The local gauge transformations are 
\bea
\delta \Xi^I &=&   \tV \AuxA^I 
\, , \nonumber\\
\delta \tAA_\a &=&   \bDD_\a \tV + (\bar{\sg} \bDD)_\a V 
\, , \nonumber\\
\delta \AuxA^I &=& 0 
\, , \label{LocalGaugeTransfSuper}\\
\delta \tBB_\a^I &=&  
    (\bar{\sg} \bDD)_\a U^I + \tV \bDD_\a \Xi^I 
 - V(\bar{\sg} \bDD)_\a \Xi^I - \tW_\a \AuxA^I
\, , \nonumber\\
\delta \tCC &=& \bDD^\a \tW_\a + \bDD^\a \tV \tAA_\a 
                - \tV \bDD^\a \tAA_\a
\, , \nonumber
\eea
where the parameters $V(z)$ and $\tV(z)$ describe 
super-${\rm U}(1)_{\rm V}$ and super-${\rm U}(1)_{\rm A}$ 
gauge transformations, respectively. 
The parameters $U^I(z)$ and $\tW_\a(z) = (\bar{\sg}W(z))_\a$ 
are related with 
the symmetry of supersymmetric generalized Chern-Simons term
whose property will be discussed in next subsection.
The global transformations are 
\bea
\delta \Xi^I &=& \omega^I{}_J \Xi^J  +  a^I
\, , \nonumber\\
\delta \tAA_\a &=&   r \tAA_\a + \sum_{i=1}^{4g} \alpha_i H^{(i)}_\a
\, , \nonumber\\
\delta \AuxA^I &=& - \, r \AuxA^I + \omega^I{}_J \AuxA^J
\, , \label{GlobalGaugeTransfSuper}\\
\delta \tBB_\a^I &=& 
r \tBB_\a^I + \omega^I{}_J \tBB_\a^J 
+ \sum_{i=1}^{4g} ( \beta_i^I + \alpha_i \Xi^I ) H^{(i)}_\a
\, , \nonumber\\
\delta \tCC &=& 2r \tCC 
\, , \nonumber
\eea
where 
$\omega_{IJ} = - \omega_{JI}$, $a^I$, $r$, $\alpha_i$ and $\beta_i^I$ 
are all constant parameters. 
Poincar\'e symmetry is ISO($D-1$,1) 
as the same as that of the bosonic model 
in subsection \ref{sec:bosonic_model}. 
$H^{(i)}_\a(z)$ 
are harmonic functions on 2D superspace which satisfy 
$\bDD H^{(i)} = \bDD \bar{\sg} H^{(i)} = 0$ 
($i=1,2,\ldots,4g$; $g=$ genus of 2D space--time), 
i.e.\ 
$H^{(i)}_\a = i(\sg^m \th)_\a h^{(i)}_m$ 
with $\prt^m h^{(i)}_m = \ep^{mn} \prt_m h^{(i)}_n = 0$ 
($i=1,2,\ldots,2g$) 
and 
$H^{(i)}_\a = i \hat{\eta}^{(i)}_\a$ 
with 
$\sg^m \prt_m \hat{\eta}^{(i)} = 0$ 
($i=2g+1,2g+2,\ldots,4g$). 
It should be noted that 
the local gauge transformation \rf{LocalGaugeTransfSuper} 
has on-shell invariance under the following 
local gauge transformation, 
\bea
\delta'  V     &=& 0
\, , \nonumber\\
\delta' \tV    &=& 0
\, , \label{LocalGaugeTransfSuper2}\\
\delta' U^I    &=& W' \AuxA^I
\, , \nonumber\\
\delta' \tW_\a &=& (\bar{\sg} \bDD)_\a W' 
\, . \nonumber
\eea
The Lagrangian is also invariant 
under the following two discrete transformations:
\bea
&&\hbox{i)}  \hspace{12pt}  \hbox{
           $\Xi^I \rightarrow -\Xi^I$, 
           $\AuxA^I \rightarrow -\AuxA^I$,
           $\tBB_\a^I \rightarrow -\tBB_\a^I$,
           otherwise unchanged,}  
\label{DiscreteTransfSuper1}\\
&&\hbox{ii)}  \hspace{8pt}  \hbox{
           $\tAA_\a \rightarrow -\tAA_\a$, 
           $\AuxA^I \rightarrow -\AuxA^I$,
           $\tBB_\a^I \rightarrow -\tBB_\a^I$,
           otherwise unchanged.} 
\label{DiscreteTransfSuper2}
\eea
Both discrete transformations are 
a kind of parity transformations of background space--time 
as the same as those of the bosonic case 
\rf{DiscreteTransfBosonic1} and \rf{DiscreteTransfBosonic2}, 
respectively. 

In the following, we give the component expression of 
fields and parameters. 
The superfields are expressed as 
\bea
\Xi^I   &=& \xi^I + i\th \l^I + \frac{i}{2} \th\th F^I
\, ,  \nonumber\\
\tAA_\a &=& i \hAy_\a + i\th_\a \tAz + i(\bar{\sg} \th)_\a \Az + 
            i(\sg^m \th)_\a \tA_m + 
            \th\th ( \Ay - \half \sg^m \prt_m \hAy )_\a
\, ,  \nonumber\\
\AuxA^I &=& \auxA^I + i\th \auyA^I + \frac{i}{2} \th\th \auzA^I
\, ,  \label{SField1}\\
\tBB_\a^I &=& i \hBy_\a^I + i\th_\a \tBz^I + i(\bar{\sg} \th)_\a \Bz^I + 
              i(\sg^m \th)_\a \tB_m^I + 
              \th\th ( \By^I - \half \sg^m \prt_m \hBy^I )_\a
\, ,  \nonumber\\
\tCC  &=& -2i ( \Cz + i\th \Cy + \frac{i}{2} \th\th \tC )
\, .  \nonumber
\eea
The gauge parameters are also expressed as 
\bea
V     &=& v + i\th \mu + \frac{i}{2} \th\th M
\, ,  \nonumber\\
\tV   &=& \tv + i\th \tmu + \frac{i}{2} \th\th \tM
\, ,  \label{SField2}\\
U^I   &=& u^I + i\th \nu^I + \frac{i}{2} \th\th N^I
\, ,  \nonumber\\
\tW_\a  &=& i \htau_\a + i\th_\a \tf + i(\bar{\sg} \th)_\a f + 
            i(\sg^m \th)_\a \tw_m + 
            \th\th ( \tau - \half \sg^m \prt_m \htau )_\a
\, .  \nonumber
\eea
The integration of Lagrangian \rf{LagrangianSuper} with respect to 
$\dd^2 \th$ gives 
\bea
\int\!\! \dd^2 \th \, \cL 
&=&  
- \, \half \prt^m \xi^I \prt_m \xi_I 
- \frac{i}{2} \l^I \sg^m \prt_m \l_I 
+ \half F^I F_I
\nonumber\\
&&
+ \, ( \tA^m \prt_m \xi^I + i \Ay \l^I - \tAz F^I ) \auxA_I
\nonumber\\
&&
+ \, \frac{i}{2}\hAy ( \sg^m \prt_m \xi^I + F^I ) \auyA_I
- \frac{i}{2}\hAy ( \sg^m \prt_m \auxA_I + \auzA_I ) \l^I
\label{LagrangianSuper2}\\
&&
+ \, \frac{i}{2} \l^I ( \tAz + \bar{\sg} \Az + \sg^m \tA_m )\auyA_I
\nonumber\\
&&
+ \, \tB^{mI} \prt_m \auxA_I + i \By^I \auyA_I - \tBz^I \auzA_I
\nonumber\\
&&
+ \, i \Cy \auyA^I \auxA_I  - \Cz \auzA^I \auxA_I
- \half \tC \auxA^I \auxA_I + \frac{i}{2} \Cz \auyA^I \auyA_I
\, .
\nonumber
\eea
The gauge transformation for each component is 
\bea
\delta \xi^I &=& \tv \auxA^I
\, ,  \nonumber\\
\delta \l^I_\a &=& \tv \auyA^I_\a + \mu'_\a \auxA^I
\, ,  \nonumber\\
\delta F^I  &=& \tv \auzA^I - i \tmu \auyA^I + \tM \auxA^I
\, ,  \nonumber\\
\delta \hAy_\a &=& (\tmu + \bar{\sg} \mu)_\a
\, ,  \nonumber\\
\delta \Az &=& M
\, ,  \nonumber\\
\delta \tAz &=& \tM
\, ,  \nonumber\\
\delta \tA_m &=& \prt_m \tv + \ep_m{}^n \prt_n v
\, ,  \nonumber\\
\delta \Ay_\a &=& (\sg^m \prt_m \tmu)_\a
\, ,  \nonumber\\
\delta \auxA^I &=& 0
\, ,  \nonumber\\
\delta \auyA^I_\a &=& 0
\, ,  \nonumber\\
\delta \auzA^I  &=& 0
\, ,  \label{STransf1}\\
\delta \hBy^I_\a &=& 
               (\bar{\sg} \nu^I)_\a 
               + \Big( (\tv - \bar{\sg} v) \l^I \Big)_\a 
               - \htau_\a \auxA^I
\, ,  \nonumber\\
\delta \Bz^I &=& N^I - v F^I 
              - \frac{i}{2} ( \tmu \bar{\sg} - \mu ) \l^I 
              - f \auxA^I - \frac{i}{2} \htau \bar{\sg} \auyA^I
\, ,  \nonumber\\
\delta \tBz^I &=& 
               \tv F^I 
               - \frac{i}{2} ( \tmu - \mu \bar{\sg} ) \l^I 
               - \tf \auxA^I + \frac{i}{2} \htau \auyA^I 
\, ,  \nonumber\\
\delta \tB^I_m &=& 
                \ep_m{}^n \prt_n u^I
                + \tv \prt_m \xi^I - v \, \ep_m{}^n \prt_n \xi^I
                - \frac{i}{2} ( \tmu + \mu \bar{\sg} ) \sg_m \l^I
\nonumber\\
           && - \, \tw_m \auxA^I - \frac{i}{2} \htau \sg_m \auyA^I
\, ,  \nonumber\\
\delta \By^I_\a &=& 
              \tv (\sg^m \prt_m \l^I)_\a 
          + \half \Big( \sg^m \prt_m (\tv - \bar{\sg} v) \l^I \Big)_\a
          - \half \Big( ( \tM - \bar{\sg} M ) \l^I \Big)_\a
\nonumber\\
           && + \, \half \Big( \sg^m (\tmu + \bar{\sg} \mu) \Big)_\a 
                   \prt_m \xi^I
              + \half (\tmu - \bar{\sg} \mu)_\a F^I 
\nonumber\\
           && - \, \tau_\a \auxA^I 
              - \half \Big( (\tf + \bar{\sg} f + \sg^m \tw_m) \auyA^I \Big)_\a
              + \half \Big( (\auzA^I - \prt_m\auxA^I \sg^m) \htau \Big)_\a
\, ,  \nonumber\\
\delta \Cz &=& \tf - \tv \tAz - \frac{i}{2} \tmu \hAy
\, ,  \nonumber\\
\delta \Cy_\a &=& \tau_\a - \tv \Ay_\a 
   - \half \Big( ( 3 \tAz - \bar{\sg} \Az - \sg^m \tA_m ) \tmu \Big)_\a
\nonumber\\&&
   - \, \half \Big( ( \tM - \prt_m \tv \sg^m ) \hAy \Big)_\a
\, ,  \nonumber\\
\delta \tC &=& \prt_m \tw^m 
            + \prt_m \tv \tA^m - \tv \prt_m \tA^m 
            + 2 i \tmu \Ay - \frac{i}{2} \prt_m (\tmu \sg^m \hAy) 
            - 2\tM \tAz
\, .  \nonumber
\eea
Now, let us redefine some of the fields and the parameters as follows: 
\bea
&& \tB^I_m - \frac{i}{2} \hAy \sg_m \l^I
\ \rightarrow \   \tB^I_m
\, ,  \nonumber\\
&& \By^I_\a + \half \Big( ( F^I - \prt_m \xi^I \sg^m ) \hAy \Big)_\a
\nonumber\\ &&\phantom{\By^I_\a}
  + \half \Big( ( \tAz - \bar{\sg} \Az - \sg^m \tA_m ) \l^I \Big)_\a
  + \Cy_\a \auxA^I + \half \auyA^I_\a \Cz
\ \rightarrow \   \By^I_\a
\, ,  \nonumber\\
&& \tBz^I + \frac{i}{2} \hAy \l^I + \Cz \auxA^I
\ \rightarrow \   \tBz^I
\, ,  \label{FieldRedefineSuper}\\
&& F^I - \tAz \auxA^I
\ \rightarrow \   F^I
\, ,  \nonumber\\
&& \tC + \tAz^2
\ \rightarrow \   \tC
\, ,  \nonumber
\eea
for the fields and
\bea
&& (\mu + \bar{\sg} \tmu)_\a
\ \rightarrow \   \mu_\a
\, ,  \nonumber\\
&& \nu^I_\a - \Big( (v - \bar{\sg} \tv) \l^I \Big)_\a - ( \bar{\sg} \htau )_\a \auxA^I
\ \rightarrow \   \nu^I_\a
\, ,  \nonumber\\
&& N^I - v F^I + \frac{i}{2} ( \mu - \tmu \bar{\sg} ) \l^I 
       - f \auxA^I - \frac{i}{2} \htau \bar{\sg} \auyA^I
\ \rightarrow \   N^I
\, ,  \nonumber\\
&& \htau_\a + \hAy_\a \tv
\ \rightarrow \   \htau_\a
\, ,  \label{ParameterRedefineSuper}\\
&& f + \tv \Az - \frac{i}{2} \tmu \bar{\sg} \hAy
\ \rightarrow \   f
\, ,  \nonumber\\
&& \tw_m - \frac{i}{2} \tmu \sg_m \hAy
\ \rightarrow \   \tw_m
\, ,  \nonumber\\
&& \tau_\a - \tv \Ay_\a
     - \half \Big( ( 3 \tAz - \bar{\sg} \Az - \sg_m \tA_m ) \tmu \Big)_\a
\nonumber\\ &&\phantom{\tau_\a - \tv \Ay_\a}
     - \half \Big( ( \tM - \prt_m \tv \sg^m ) \hAy \Big)_\a
\ \rightarrow \   \tau_\a
\, ,  \nonumber
\eea
for the parameters. 
Then, the Lagrangian \rf{LagrangianSuper2} becomes a simple form, 
\bea
\int\!\! \dd^2 \th \, \cL 
&=&  
- \, \half \prt^m \xi^I \prt_m \xi_I 
- \frac{i}{2} \l^I \sg^m \prt_m \l_I 
+ \half F^I F_I
\nonumber\\
&&
+ \, ( \tA^m \prt_m \xi^I + i \Ay \l^I ) \auxA_I
\label{LagrangianSuper3}\\
&&
+ \, \tB^{mI} \prt_m \auxA_I
+ i \By^I \auyA_I
- \tBz^I \auzA_I
- \, \half \tC \auxA^I \auxA_I      
\, .
\nonumber
\eea
The gauge transformation \rf{STransf1} also becomes 
simple as 
\bea
\delta \xi^I &=& \tv \auxA^I
\, ,  \nonumber\\
\delta \l^I_\a &=& \tv \auyA^I_\a + \mu'_\a \auxA^I
\, ,  \nonumber\\
\delta F^I  &=& \tv \auzA^I - i \tmu \auyA^I
\, ,  \nonumber\\
\delta \tA_m &=& \prt_m \tv + \ep_m{}^n \prt_n v
\, ,  \nonumber\\
\delta \Ay_\a &=& (\sg^m \prt_m \tmu)_\a
\, ,  \nonumber\\
\delta \auxA^I &=& 0
\, ,  \nonumber\\
\delta \auyA^I_\a &=& 0
\, ,  \nonumber\\
\delta \auzA^I  &=& 0
\, ,  \label{STransf2}\\
\delta \tBz^I &=& 
                  \tv F^I + \frac{i}{2} \htau \auyA^I 
\, ,  \nonumber\\
\delta \tB^I_m &=& 
                   \ep_m{}^n \prt_n u^I
                   + \tv \prt_m \xi^I - v \, \ep_m{}^n \prt_n \xi^I
                   - i \tmu \sg_m \l^I
                   - \tw_m \auxA^I - \frac{i}{2} \htau \sg_m \auyA^I
\, ,  \nonumber\\
\delta \By^I_\a &=& 
                    \tv (\sg^m \prt_m \l^I)_\a
                    - \tv \Ay_\a \auxA^I + \mu'_\a F^I
                    - \half f (\bar{\sg} \auyA^I)_\a
\nonumber\\
                 && - \, \half (\tw_m + \tv \tA_m) (\sg^m \auyA^I)_\a 
                    + \half \Big( (\auzA^I - \prt_m\auxA^I \sg^m) 
                                  \htau \Big)_\a
\, ,  \nonumber\\
\delta \tC &=& \prt_m \tw^m 
               + \prt_m \tv \tA^m - \tv \prt_m \tA^m
               + 2 i \tmu \Ay
\, ,  \nonumber
\eea
and
\bea
\delta \hAy_\a &=& (\bar{\sg} \mu)_\a
\, ,  \nonumber\\
\delta \Az &=& M
\, ,  \nonumber\\
\delta \tAz &=& \tM
\, ,  \nonumber\\
\delta \hBy^I_\a &=& 
                 (\bar{\sg} \nu^I)_\a
\, ,  \label{STransf3}\\
\delta \Bz^I &=& N^I
\, ,  \nonumber\\
\delta \Cz &=& \tf
\, ,  \nonumber\\
\delta \Cy_\a &=& \tau_\a
\, .  \nonumber
\eea
Note that 
no gauge fixing procedures are done at this stage, 
so the Lagrangian \rf{LagrangianSuper3} still has supersymmetry. 
The fields 
$\hAy_\a$, $\Az$, $\tAz$, 
$\hBy^I_\a$, $\Bz^I$, $\Cz$, $\Cy_\a$  
do not exist in the Lagrangian \rf{LagrangianSuper3} 
and can be gauged away easily by using \rf{STransf3}. 
The field redefinition \rf{FieldRedefineSuper} 
has decoupled these fields. 

\subsection{Supersymmetric generalized Chern-Simons action}
\label{sec:GCS_supersymmetric_action}

In this subsection we will show that 
a part of the Lagrangian \rf{LagrangianSuper}, 
\beq{LagrangianGCSsuper}
\cL_{\rm sGCS} \ = \  
\tBB^{\a I} \bDD_\a \AuxA_I
- \half \tCC \AuxA^I \AuxA_I
\, ,
\eeq
is a kind of the supersymmetric version of 
the generalized Chern-Simons Lagrangian.

We here propose 
the supersymmetric generalized Chern-Simons theory 
in two dimensions. 
The Lagrangian of 
the supersymmetric generalized Chern-Simons theory is 
\beq{LagrangianGCSsuper2}
\cL_{\rm sGCS} \ = \  
\Htr \Bigl\{ \AuxA ( \bDD \bar{\sg} \Pi + \Pi \bar{\sg} \Pi ) 
             + \AuxA^2 \L \Bigr\}
\, ,
\eeq
where 
\beq{GCSrepsuper1}
\AuxA    \ = \  \half \Sigma_{\hat \a} \AuxA^{\hat \a}  \, ,
~~~~~~
\Pi_\a   \ = \  \half T_{\hat a} \Pi_\a^{\hat a}        \, ,
~~~~~~
\L       \ = \  \half \Sigma'_{\hat \a'} \L^{\hat \a'}   \, .
\eeq
The Lagrangian \rf{LagrangianGCSsuper2} is invariant 
under the local gauge transformation:
\bea
\delta \AuxA   &=& [ \AuxA , U ] 
\, ,
\nonumber\\
\delta \Pi_\a  &=& \bDD_\a U + [ \Pi_\a , U ] - \{ \AuxA , \Upsilon_\a \}
\, ,
\label{GCStransfsuper1}\\
\delta \L      &=& \bDD \bar{\sg} \Upsilon 
                   + \Pi \bar{\sg} \Upsilon + \Upsilon \bar{\sg} \Pi 
                   + [ \L , U ] - [ \AuxA , Q ]
\, ,
\nonumber
\eea
where 
\beq{GCSrepsuper2}
U      \ = \ \half T_{\hat a} U^{\hat a}                     \, ,
~~~~~~
\Upsilon_\a  \ = \ \half \Sigma'_{\hat \a'} W_\a^{\hat \a'}  \, ,
~~~~~~
Q  \ = \  \half T'_{\hat a'} Q^{\hat a'}                     \, .
\eeq
Note that 
$\Sigma_{\hat \a}$ and $\Sigma'_{\hat \a'}$ 
are not Grassmannian. 

By the following identification, 
\bea
\bDD_1  &=&   \dd x^0 \prt_0 
\, , ~~~~~~~
\bDD_2 \ = \  \dd x^1 \prt_1 
\, ,
\nonumber\\
\Pi_1  &=&    \dd x^0 B_0 
\, , \hspace{25pt}
\Pi_2 \ = \   \dd x^1 B_1 
\, ,
\label{BosonicSuperRelation}\\
\L  &=&  \L_{12}   \ = \   \dd x^0 \!\!\wedge\! \dd x^1 C_{01} 
\, ,
\nonumber
\eea
we find that 
the supersymmetric version of the generalized Chern-Simons Lagrangian 
proposed above is equivalent to 
the generalized Chern-Simons Lagrangian. 
Namely, 
the algebraic structure of both theories are the same.
In the supersymmetric generalized Chern-Simons theory 
the matrix $\bar{\sg}$ plays a role of the wedge product $\wedge$ 
in the generalized Chern-Simons theory.
The supercovariant derivative $\bDD$ corresponds to 
the external derivative $\dd$, 
so the property $\bDD \bar{\sg} \bDD=0$ corresponds to 
$\dd \!\wedge\! \dd = 0$. 
Thus, one can expect that 
$2^{D/2}$-dimensional generalized Chern-Simons theory 
will be deeply related with 
$D$-dimensional supersymmetric 
generalized Chern-Simons theory. 

In order to obtain \rf{LagrangianGCSsuper} 
from the general expression \rf{LagrangianGCSsuper2}, 
one chooses the representation \rf{GCSrepresentation}, i.e.\ 
\bea
\AuxA  &=& \half \Gamma_I \AuxA^I + 
           \quarter \sg_+ [ \Gamma_I , \Gamma_J ] 
           \AuxA^{\prime IJ}
\, ,
\nonumber\\
\Pi_\a &=& \half \sg_+ \Gamma_I \Pi^I_\a
\, ,
\label{GCSrepsuper3}\\
 \L    &=& \half \sg_+ \tCC 
\, ,
\nonumber
\eea
and 
\bea
U &=& \half \sg_+ \Gamma_I U^I 
\, ,
\nonumber\\
\Upsilon_\a &=& \half \sg_+ W_\a
\, ,
\label{GCSrepsuper4}\\
Q   &=& \half \sg_+ {\tilde Q}
\, ,
\nonumber
\eea
where 
$\Pi^I_\a = (\bar{\sg} \tBB^I)_\a$ and 
$W_\a = (\bar{\sg} \tW)_\a$. 
As the same as in the bosonic case, 
$\AuxA^{\prime IJ}$ and $\tilde Q$ are decoupled.

\subsection{The gauge fixing of {\UvUa} supersymmetric model}
\label{sec:gaugefixing_supersymmetric_model}

Using the gauge transformations \rf{STransf3}, 
one easily finds that 
the gauge parameters 
$\mu_\a$, $M$, $\tM$, 
$\nu^I_\a$, $N^I$, $\tf$, $\tau_\a$  
gauge away the fields 
$\hAy_\a$, $\Az$, $\tAz$, 
$\hBy^I_\a$, $\Bz^I$, $\Cz$, $\Cy_\a$, respectively.  
By this gauge fixing 
the Lagrangian \rf{LagrangianSuper3} is unchanged 
because the fields 
$\hAy_\a$, $\Az$, $\tAz$, 
$\hBy^I_\a$, $\Bz^I$, $\Cz$, $\Cy_\a$  
do not exist in the Lagrangian. 
The gauge transformation \rf{STransf2} is also unchanged 
by the same reason. 

Finally, 
let us perform the integration of $F^I$, $\By^I_\a$, and $\tBz^I$. 
This procedure gives simple equations of motion, 
$F^I = 0$, $\auyA^I_\a = 0$, and $\auzA^I = 0$, 
because they are decoupled. 
Then, we find 
\bea
\int\!\! \dd^2 \th \, \cL 
&=&  
- \, \half \prt^m \xi^I \prt_m \xi_I 
- \frac{i}{2} \l^I \sg^m \prt_m \l_I 
\label{LagrangianSuper4}\\
&&
+ \, ( \tA^m \prt_m \xi^I + i \Ay \l^I ) \, \auxA_I
+ \tB^{mI} \prt_m \auxA_I
- \half \tC \auxA^I \auxA_I                 
\, .  \nonumber
\eea
The gauge transformation for each component is 
\bea
\delta \xi^I &=& \tv \auxA^I
\, ,  \nonumber\\
\delta \l^I_\a &=& \mu'_\a \auxA^I
\, ,  \nonumber\\
\delta \tA_m &=& \prt_m \tv + \ep_m{}^n \prt_n v
\, ,  \nonumber\\
\delta \Ay_\a &=& (\sg^m \prt_m \tmu)_\a
\, ,  \label{STransf4}\\
\delta \auxA^I &=& 0
\, ,  \nonumber\\
\delta \tB^I_m &=& 
                   \ep_m{}^n \prt_n u^I 
                   + \tv \prt_m \xi^I - v \, \ep_m{}^n \prt_n \xi^I
                   - i \tmu \sg_m \l^I - \tw_m \auxA^I
\, ,  \nonumber\\
\delta \tC &=& \prt_m \tw^m 
               + \prt_m \tv \tA^m - \tv \prt_m \tA^m
               + 2 i \tmu \Ay 
\, .  \nonumber
\eea
The Lagrangian \rf{LagrangianSuper4} does not have 
off-shell supersymmetry any more, 
but is still off-shell invariant 
under the gauge transformation \rf{STransf4}. 
The remaining gauge parameters are 
$v$, $\tv$, $\mu'_\a$, $u^I$, and $\tw_m$.
It should be noted that 
the generalized Chern-Simons term in \rf{LagrangianSuper4}
does not have the supersymmetric counterparts. 
This means that the supersymmetry does not lead us to 
a new topological feature 
concerning the generalized Chern-Simons term 
appeared in this model.

As the same as in the {\UvUa} bosonic model, 
using the global Poincar\'e ISO($D-1$,1) symmetry 
and the internal symmetry, 
the solution \rf{auxiliaryfieldSolution} is achieved 
from the equations of motion without loss of generality. 
Then, the Lagrangian \rf{LagrangianSuper4} becomes 
\beq{LagrangianLCsuper}
\cL \ = \  
- \, \half \, \prt^m \xi^I \prt_m \xi_I 
- \frac{i}{2} \l^I \sg^m \prt_m \l_I 
+ \frac{1}{\sqrt{\pi}} 
( \tA^m \prt_m \xi^{\widehat{-}} + i \Ay \l^{\widehat{-}} )
\, ,
\eeq
which is the supersymmetric version of \rf{LagrangianLC}. 
In the case of the solution \rf{auxiliaryfieldSolution}, 
the gauge transformation 
$\delta \xi^I$ and $\delta \l^I_\a$ in \rf{STransf4} 
make us possible to choose 
the light-cone-like gauge fixing condition, 
\beq{LightConeSuper}
\xi^{\widehat{+}}  \ = \  0 
\, , ~~~~~~
\l^{\widehat{+}}_\a = 0
\, .
\eeq
The gauge fields $\tA^m$ and $\Ay$ play roles of 
Lagrange multipliers. 

\subsection{{\UvUa} Neveu-Schwarz-Ramond type Superstring}
\label{sec:NSR-superstring}

In this subsection we consider 
the {\UvUa} supersymmetric model coupled to 2D supergravity, 
which is equivalent to 
the Neveu-Schwarz-Ramond type \cite{SuperString} of 
{\UvUa} superstring model. 

The dynamical variables of 2D supergravity are 
a vielbein $E^A = \dd z^M E_M{}^A(z)$ and 
a connection $\Omega_A{}^B = \dd z^M \Omega_M(z) \ep_A{}^B$, 
where $A = (a, \a)$ ($a = +$, $-$; $\a = 1$, $2$) 
is a tangent space index, 
and 
\beq{generatorSO2}
\ep_A{}^B  \ = \ 
    \left(\begin{array}{cc}
             \ep_a{}^b &            0              \\
                  0    & \frac{1}{2} (\bar{\sg})_\alpha{}^\beta
    \end{array}\right) 
\eeq
is the generator of SO(1,1) tangent group.
The kinematical constraints on the torsion 
$T^A = \DD E^A = - \frac{1}{2} E^C E^B T_{BC}{}^A(z)$ 
are \cite{howe}
\beq{TorsionlessCondition2DN1}
T_{\beta\gamma}{}^a(z)  =  2 i (\sg^a)_{\beta\gamma}
\, , ~~~~~~
T_{bc}{}^a(z)  =  T_{\beta\gamma}{}^\alpha(z)  =  0
\, .
\eeq
Other torsion components are determined 
from the constraints \rf{TorsionlessCondition2DN1}.

The Lagrangian for 
the {\UvUa} supersymmetric model coupled to supergravity is 
\beq{LagrangianSuperGravity}
\cL \ = \  E \, \Big(
- \half \DD^\a \Xi^I \DD_\a \Xi_I 
+ \tAA^\a \AuxA_I \DD_\alpha \Xi^I 
+ \tBB^{\a I} \DD_\a \AuxA_I
- \half \tCC \AuxA^I \AuxA_I  \Big)
\, ,
\eeq
where 
$E(z) = \sdet E_M{}^A(z) 
      = \det E_m{}^a(z) \det E_\alpha{}^\mu(z)$. 
The background metric $\eta_{IJ}$ is \rf{FlatmetricDefinitionGravity}. 
This Lagrangian is invariant 
under the local super {\UvUa} transformations, 
the supergeneral coordinate transformations,
the local Lorentz transformations, and 
the super-Weyl scaling as follows:
\bea
\delta \Xi^I &=&   \tV \AuxA^I 
\, , \nonumber\\
\delta \tAA_\a &=&  \DD_\a \tV + (\bar{\sg} \DD)_\a V 
\, , \nonumber\\
\delta \AuxA^I &=& 0 
\, , \nonumber\\
\delta \tBB_\a^I &=&  
     (\bar{\sg} \DD)_\a U^I + \tV \DD_\a \Xi^I 
 - V (\bar{\sg} \DD)_\a \Xi^I - \tW_\a \AuxA^I
\, , \label{LocalGaugeTransfSuperGravity}\\
\delta \tCC  &=&  \DD^\a \tW_\a + \DD^\a \tV \tAA_\a - \tV \DD^\a \tAA_\a
\, , \nonumber\\
\delta E_M{}^A   &=&  0
\, , \nonumber\\
\delta \Omega_M  &=&  0
\, , \nonumber
\eea
and
\bea
\delta \Xi^I &=&   K^N \prt_N \Xi^I
\, , \nonumber\\
\delta \tAA_\a &=&   K^N \prt_N \tAA_\a 
                   - \frac{1}{2} L \, (\bar{\sg})_\a{}^\b \tAA_\b 
                   + \frac{1}{2} S \tAA_\a
\, , \nonumber\\
\delta \AuxA^I &=& K^N \prt_N \AuxA^I
\, , \nonumber\\
\delta \tBB_\a^I &=& K^N \prt_N \tBB_\a^I 
                   - \frac{1}{2} L \, (\bar{\sg})_\a{}^\b \tBB_\b^I
                   + \frac{1}{2} S \tBB_\a^I
\, , \nonumber\\
\delta \tCC &=&     K^N \prt_N \tCC + S \tCC
\, , \label{GeneralCoordinateTransfSuperGravity}\\
\delta E_M{}^a       &=&
  K^N \prt_N E_M{}^a + \prt_M K^N E_N{}^a + E_M{}^b L \, \ep_b{}^a 
   - S E_M{}^a  
\, , \nonumber\\
\delta E_M{}^\a  &=&
  K^N \prt_N E_M{}^\a + \prt_M K^N E_N{}^\a 
  + \frac{1}{2} E_M{}^\b L \, (\bar{\sg})_\b{}^\a  \nonumber\\&&
  - \, \frac{1}{2} S E_M{}^\a 
  + \frac{i}{2} E_M{}^a (\sg_a)^{\a\b} \DD_\b S
\, , \nonumber\\
\delta \Omega_M      &=&
  K^N \prt_N \Omega_M + \prt_M K^N \Omega_N + \prt_M L  \nonumber\\&&
  + \, E_M{}^a \ep_a{}^b \DD_b S 
  + E_M{}^\a (\bar{\sg})_\a{}^\b \DD_\b S
\, , \nonumber
\eea
where $K^N(z)$, $L(z)$ and $S(z)$ are local parameters for 
the supergeneral coordinate transformation, 
the local Lorentz transformations, and 
the super-Weyl transformation, respectively. 
The gauge transformation \rf{LocalGaugeTransfSuperGravity} 
has on-shell reducibility 
as the same as the gauge transformation \rf{LocalGaugeTransfSuper}.
The transformation \rf{LocalGaugeTransfSuperGravity} 
has on-shell invariance under the two-dimensional covariant form 
of the local gauge transformation \rf{LocalGaugeTransfSuper2}. 
The two-dimensional covariant form is obtained 
by replacing $\bDD$ by $\DD$. 
The Lagrangian \rf{LagrangianSuperGravity} is also invariant 
under the global transformations \rf{GlobalGaugeTransfSuper} 
and the discrete transformations 
\rf{DiscreteTransfSuper1} and \rf{DiscreteTransfSuper2}, 
where $H^{(i)}_\a(z)$ 
are harmonic functions on 2D superspace which satisfy 
$\DD H^{(i)} = \DD \bar{\sg} H^{(i)} = 0$ 
($i=1,2,\ldots,4g$). 
The Poincar\'e symmetry is ISO($D-2$,2). 

At the quantum level, 
the absence of superconformal anomaly requires $D=12$. 
The {\UvUa} superstring model consists of 
one graviton ($j=2$), one gravitino ($j=3/2$), 
one photon ($j=1$), one photino ($j=1/2$), 
$D$ scalar particles ($j=0$), 
and 
$D$ spinor particles ($j=1/2$). 
Therefore, 
the total conformal charge of {\UvUa} superstring model is 
\bea
w^{(N=1)} 
&=&  2 \!\times\! (-13) + 2 \!\times\! \frac{11}{2} + 
     2 \!\times\! (-1) + 2 \!\times\! (-\frac{1}{2}) + 
     D \!\times\! (1 + \frac{1}{2})
\label{ConformalWeightSuper}\\
&=&  \frac{3}{2} ( D - 12 ) 
\, . \nonumber
\eea
The fields $\AuxA^I$, $\tBB_m^I$, and $\tCC$ 
do not contribute to the conformal charge 
because these fields come from the generalized Chern-Simons action 
which is considered to be topological. 
The cancellation of superconformal anomaly, i.e.\ 
the existence of the super-Weyl symmetry at the quantum level 
requires $w^{(N=1)} = 0$. 
Thus, we obtain $D=12$ from \rf{ConformalWeightSuper}.

\section{{\UvUa} Green-Schwarz type Superstring}
\label{sec:GS-superstring}

In this section we try to study 
the Green-Schwarz type \cite{SuperString}
of {\UvUa} superstring model. 
This model has a kind of manifest supersymmetry 
in the background target space--time. 
Thus, one needs 
not only the background space--time coordinate $\xi^I$, 
but also 12 dimensional fermionic spinor coordinates 
$\Theta^1$ and $\Theta^2$. 
Similar to the Green-Schwarz superstring model, 
we require $\Theta^1$ and $\Theta^2$ to be 
scalar fields in two-dimensional field theory
and Majorana-Weyl spinors in 12 dimensions. 
Note that it is possible to require 
both Majorana condition and Weyl condition in 12 dimensions 
at the same time owing to the existence of two time coordinates
\cite{wetterich}. 
In order to obtain the supersymmetry, 
the number of bosonic freedoms and 
that of fermionic freedoms are set to be equal. 
So, 
the gauge transformation 
$\delta \Theta^1$ and $\delta \Theta^2$ will lead us to 
the light-cone-like gauge fixing conditions, 
\beq{LightConeTheta}
\Gamma^{\widehat{+}} \Theta^1  \ = \  0  \, , ~~~~~~
\Gamma^{\widehat{+}} \Theta^2  \ = \  0  \, ,
\eeq
as the same as 
\rf{LightConeBosonic} and \rf{LightConeSuper}. 

Since the way how to covariantize the Lagrangian is the same as 
that in the bosonic type and 
that in the Neveu-Schwarz-Ramond type, 
we here give the final expression. 
The covariant Lagrangian of 
the Green-Schwarz type of {\UvUa} superstring model is\footnote{
A similar Lagrangian is obtained in \cite{Bars1}, 
but the gauge structure is different.
Not only {\UvUa} symmetry but also $\pi$-symmetry 
play essential roles in our model.
}
\bea
\cL  &=&  \sqrt{-g} \, \Big\{
- \half g^{mn} \Pi_m^I \Pi_{nI} 
\nonumber\\
&&\phantom{\sqrt{-g} \, \Big\{}
- i \, \EP^{mn} \Pi_m^I 
   ( \Theta^1 \Gamma_{IJ} \prt_n \Theta^1 - 
     \Theta^2 \Gamma_{IJ} \prt_n \Theta^2 ) \auxA^J
\label{LagrangianGSstring}\\
&&\phantom{\sqrt{-g} \, \Big\{}
+ \EP^{mn}
     \Theta^1 \Gamma^K{}_I \prt_m \Theta^1
     \Theta^2 \Gamma_{KJ}  \prt_n \Theta^2  \auxA^I \auxA^J
\nonumber\\
&&\phantom{\sqrt{-g} \, \Big\{}
+ \tA^m \auxA_I \Pi_m^I
+ \tB^{mI} \prt_m \auxA_I
- \half \tC \auxA^I \auxA_I
\Big\}
\, , \nonumber
\eea
where
\beq{Pi}
\Pi_m^I  \ = \  
\prt_m \xi^I 
 + i ( \Theta^1 \Gamma^I{}_J \prt_m \Theta^1 + 
       \Theta^2 \Gamma^I{}_J \prt_m \Theta^2) \auxA^J
\, .
\eeq
The chirality of $\Theta^1$ and $\Theta^2$ is defined as follows: 
\bea
&&
\bar{\Gamma} \Theta^1 = \pm \Theta^1 \, , ~~~~
\bar{\Gamma} \Theta^2 = \mp \Theta^2 ~~ 
\cdots\cdots ~~ \hbox{type IIA}, 
\\
&&
\bar{\Gamma} \Theta^1 = \pm \Theta^1 \, , ~~~~
\bar{\Gamma} \Theta^2 = \pm \Theta^2 ~~ 
\cdots\cdots ~~ \hbox{type IIB and type I}. 
\nonumber
\eea
The Majorana conditions are also necessary to 
$\Theta^1$ and $\Theta^2$. 

The local gauge symmetries of 
the Lagrangian \rf{LagrangianGSstring} are 
not only 
{\UvUa} gauge symmetry, 
general coordinate invariance, and Weyl symmetry, 
but also 
$\kappa$-symmetry, $\pi$-symmetry, and $\lambda$-symmetry 
defined in the following. 
The {\UvUa} gauge transformation is 
\bea
\delta \Theta^1  &=&  0
\, , \nonumber\\
\delta \Theta^2  &=&  0
\, , \nonumber\\
\delta \xi^I     &=&  \tv \auxA^I
\, , \nonumber\\
\delta \tA_m     &=&  \prt_m \tv + \EP_m{}^n \prt_n v
\, , \nonumber\\
\delta \auxA^I   &=&  0
\, , \label{UvUaGaugeSymmetry}\\
\delta \tB_m^I  &=&  
    \EP_m{}^n \prt_n u^I 
    + \tv \prt_m \xi^I 
    + 2 i \tv ( 
       P_{+m}{}^n \Theta^1 \Gamma^I{}_J \prt_n \Theta^1 + 
       P_{-m}{}^n \Theta^2 \Gamma^I{}_J \prt_n \Theta^2 ) \auxA^J
\nonumber\\
&&
- \, v \, \EP_m{}^n \prt_n \xi^I - \tw_m \auxA^I
\, , \nonumber\\
\delta \tC       &=&  \nabla\!_m \tw^m 
               + \prt_m \tv \tA^m - \tv \nabla\!_m \tA^m
\, , \nonumber\\
\delta g_{mn}    &=&  0
\, , \nonumber
\eea
where 
$P_\pm^{mn}$ are projection tensors 
defined in \rf{ProjectionOperator}. 
The general coordinate transformation and 
the Weyl transformation are 
\bea
\delta \Theta^1 &=&   k^n \prt_n \Theta^1
\, , \nonumber\\
\delta \Theta^2 &=&   k^n \prt_n \Theta^2
\, , \nonumber\\
\delta \xi^I    &=&   k^n \prt_n \xi^I
\, , \nonumber\\
\delta \tA_m    &=&   k^n \prt_n \tA_m + \prt_m k^n \tA_n
\, , \label{GeneralCoordinateTransfGSgravity}\\
\delta \auxA^I  &=&   k^n \prt_n \auxA^I
\, , \nonumber\\
\delta \tB_m^I  &=&   k^n \prt_n \tB_m^I + \prt_m k^n \tB_n^I
\, , \nonumber\\
\delta \tC      &=&   k^n \prt_n \tC + 2 s \tC
\, , \nonumber\\
\delta g_{mn} &=&  k^l \prt_l g_{mn} + \prt_m k^l g_{ln} + \prt_n k^l g_{ml}
                   - 2 s g_{mn}
\, . \nonumber
\eea
The $\kappa$-transformation and the $\pi$-transformation 
are fermionic transformations and have the forms, 
\bea
\delta \Theta^1  &=&  \Gamma_I \k^{1m} \Pi_m^I
\, , \nonumber\\
\delta \Theta^2  &=&  \Gamma_I \k^{2m} \Pi_m^I
\, , \nonumber\\
\delta \xi^I     &=&  i ( \delta \Theta^1 \Gamma^I{}_J \Theta^1 + 
                          \delta \Theta^2 \Gamma^I{}_J \Theta^2 ) \auxA^J
\, , \nonumber\\
\delta \tA^m     &=&  
- \, 4i \, (
P_+^{ml} \k^{1n} \Gamma_I \prt_l \Theta^1 + 
P_-^{ml} \k^{2n} \Gamma_I \prt_l \Theta^2 
) \Pi_n^I
\, , \nonumber\\
\delta \auxA^I    &=&  0
\, , \label{KappaSymmetry}\\
\delta \tB^{mI}   &=&  
   - \, 2i \, 
    ( \delta \Theta^1 \Gamma^I{}_J \Theta^1 P_-^{mn} + 
      \delta \Theta^2 \Gamma^I{}_J \Theta^2 P_+^{mn} ) \Pi_n^J
\nonumber\\
&&- \, \EP^{mn} ( 
  \delta \Theta^1 \Gamma^{KI} \Theta^1 
  \Theta^1 \Gamma_{KJ} \prt_n \Theta^1 
- \delta \Theta^2 \Gamma^{KI} \Theta^2 
  \Theta^2 \Gamma_{KJ} \prt_n \Theta^2 
\nonumber\\
&&\phantom{- \, \EP^{mn} (}\hspace{-14.7pt}
- \delta \Theta^1 \Gamma_{KJ} \Theta^1 
  \Theta^1 \Gamma^{KI} \prt_n \Theta^1 
+ \delta \Theta^2 \Gamma_{KJ} \Theta^2 
  \Theta^2 \Gamma^{KI} \prt_n \Theta^2 
) \auxA^J
\nonumber\\
&&+ \, i \, \tA^m ( 
  \delta \Theta^1 \Gamma^I{}_J \Theta^1 
+ \delta \Theta^2 \Gamma^I{}_J \Theta^2 ) 
\auxA^J
\, , \nonumber\\
\delta \tC   &=&  
- \, \frac{1}{6} \, \EP^{mn} \Big\{ 
\delta \Theta^1 \Gamma^{IJ} ( 
  \Theta^1 \prt_m \Theta^1 \Gamma_{IJ} \prt_n \Theta^1 
+ 2 \prt_m \Theta^1 \prt_n \Theta^1 \Gamma_{IJ} \Theta^1 ) 
\nonumber\\
&&\phantom{- \, \frac{1}{6} \, \EP^{mn} \Big\{}\hspace{-14.7pt}
- \delta \Theta^2 \Gamma^{IJ} ( 
  \Theta^2 \prt_m \Theta^2 \Gamma_{IJ} \prt_n \Theta^2 
+ 2 \prt_m \Theta^2 \prt_n \Theta^2 \Gamma_{IJ} \Theta^2 ) 
\Big\}
\, , \nonumber\\
\delta g^{mn}   &=&  
- \, 8i \, (
P_+^{ml} \k^{1n} \Gamma_I \prt_l \Theta^1 + 
P_-^{ml} \k^{2n} \Gamma_I \prt_l \Theta^2 
) \auxA^I
\, , \nonumber
\eea
and 
\bea
\delta \Theta^1  &=&  \Gamma_I \pi^1 \auxA^I
\, , \nonumber\\
\delta \Theta^2  &=&  \Gamma_I \pi^2 \auxA^I
\, , \nonumber\\
\delta \xi^I     &=&  i ( \delta \Theta^1 \Gamma^I{}_J \Theta^1 + 
                          \delta \Theta^2 \Gamma^I{}_J \Theta^2 ) \auxA^J
\, , \nonumber\\
\delta \tA^m     &=&  
4i \, (
P_+^{mn} \pi^1 \Gamma_I \prt_n \Theta^1 + 
P_-^{mn} \pi^2 \Gamma_I \prt_n \Theta^2 
) \auxA^I
\, , \nonumber\\
\delta \auxA^I    &=&  0
\, , \label{PiSymmetry}\\
\delta \tB^{mI}   &=&  
   - \, 2i \, 
    ( \delta \Theta^1 \Gamma^I{}_J \Theta^1 P_-^{mn} + 
      \delta \Theta^2 \Gamma^I{}_J \Theta^2 P_+^{mn} ) \Pi_n^J
\nonumber\\
&&- \, \EP^{mn} ( 
  \delta \Theta^1 \Gamma^{KI} \Theta^1 
  \Theta^1 \Gamma_{KJ} \prt_n \Theta^1 
- \delta \Theta^2 \Gamma^{KI} \Theta^2 
  \Theta^2 \Gamma_{KJ} \prt_n \Theta^2 
\nonumber\\
&&\phantom{- \, \EP^{mn} (}\hspace{-14.7pt}
- \delta \Theta^1 \Gamma_{KJ} \Theta^1 
  \Theta^1 \Gamma^{KI} \prt_n \Theta^1 
+ \delta \Theta^2 \Gamma_{KJ} \Theta^2 
  \Theta^2 \Gamma^{KI} \prt_n \Theta^2 
) \auxA^J
\nonumber\\
&&+ \, i \, \tA^m ( 
  \delta \Theta^1 \Gamma^I{}_J \Theta^1 
+ \delta \Theta^2 \Gamma^I{}_J \Theta^2 ) 
\auxA^J
\, , \nonumber\\
\delta \tC    &=&  
- \, \frac{1}{6} \, \EP^{mn} \Big\{ 
\delta \Theta^1 \Gamma^{IJ} ( 
  \Theta^1 \prt_m \Theta^1 \Gamma_{IJ} \prt_n \Theta^1 
+ 2 \prt_m \Theta^1 \prt_n \Theta^1 \Gamma_{IJ} \Theta^1 ) 
\nonumber\\
&&\phantom{- \, \frac{1}{6} \, \EP^{mn} \Big\{}\hspace{-14.7pt}
- \delta \Theta^2 \Gamma^{IJ} ( 
  \Theta^2 \prt_m \Theta^2 \Gamma_{IJ} \prt_n \Theta^2 
+ 2 \prt_m \Theta^2 \prt_n \Theta^2 \Gamma_{IJ} \Theta^2 ) 
\Big\}
\, , \nonumber\\
&&+ \, 8 i \, \Pi_m^I (
  P_+^{mn} \pi^1 \Gamma_I \prt_n \Theta^1 
+ P_-^{mn} \pi^2 \Gamma_I \prt_n \Theta^2 )
\nonumber\\
\delta g^{mn}    &=&  0
\, . \nonumber
\eea
The $\pi$-symmetry is a new local supersymmetry 
which does not exist in the usual Green-Schwarz superstring. 
The $\lambda$-transformation is a bosonic transformation 
and has the form, 
\bea
\delta \Theta^1  &=&  \lambda^{1m} \prt_m \Theta^1
\, , \nonumber\\
\delta \Theta^2  &=&  \lambda^{2m} \prt_m \Theta^2
\, , \nonumber\\
\delta \xi^I     &=&  i ( \delta \Theta^1 \Gamma^I{}_J \Theta^1 + 
                          \delta \Theta^2 \Gamma^I{}_J \Theta^2 ) \auxA^J
\, , \nonumber\\
\delta \tA^m     &=&  0
\, , \nonumber\\
\delta \auxA^I   &=&  0
\, , \label{LambdaSymmetry}\\
\delta \tB^{mI}  &=&  
   - \, 2i \, 
    ( \delta \Theta^1 \Gamma^I{}_J \Theta^1 P_-^{mn} + 
      \delta \Theta^2 \Gamma^I{}_J \Theta^2 P_+^{mn} ) \Pi_n^J
\nonumber\\
&&- \, \EP^{mn} ( 
  \delta \Theta^1 \Gamma^{KI} \Theta^1 
  \Theta^1 \Gamma_{KJ} \prt_n \Theta^1 
- \delta \Theta^2 \Gamma^{KI} \Theta^2 
  \Theta^2 \Gamma_{KJ} \prt_n \Theta^2 
\nonumber\\
&&\phantom{- \, \EP^{mn} (}\hspace{-14.7pt}
- \delta \Theta^1 \Gamma_{KJ} \Theta^1 
  \Theta^1 \Gamma^{KI} \prt_n \Theta^1 
+ \delta \Theta^2 \Gamma_{KJ} \Theta^2 
  \Theta^2 \Gamma^{KI} \prt_n \Theta^2 
) \auxA^J
\nonumber\\
&&+ \, i \, \tA^m ( 
  \delta \Theta^1 \Gamma^I{}_J \Theta^1 
+ \delta \Theta^2 \Gamma^I{}_J \Theta^2 ) 
\auxA^J
\, , \nonumber\\
\delta \tC    &=&  
- \, \frac{1}{6} \, \EP^{mn} \Big\{ 
\delta \Theta^1 \Gamma^{IJ} ( 
  \Theta^1 \prt_m \Theta^1 \Gamma_{IJ} \prt_n \Theta^1 
+ 2 \prt_m \Theta^1 \prt_n \Theta^1 \Gamma_{IJ} \Theta^1 ) 
\nonumber\\
&&\phantom{- \, \frac{1}{6} \, \EP^{mn} \Big\{}\hspace{-14.5pt}
- \delta \Theta^2 \Gamma^{IJ} ( 
  \Theta^2 \prt_m \Theta^2 \Gamma_{IJ} \prt_n \Theta^2 
+ 2 \prt_m \Theta^2 \prt_n \Theta^2 \Gamma_{IJ} \Theta^2 ) 
\Big\}
\, , \nonumber\\
\delta g^{mn}   &=&  0
\, . \nonumber
\eea
Note that $\delta g = 0$ under the 
$\kappa$-, $\pi$-, and $\lambda$-transformations. 
The local parameters 
$\kappa^{1m}$, $\kappa^{2m}$, $\pi^1$, and $\pi^2$ 
are 12-dimensional Majorana-Weyl spinors and 
have the opposite chirality to 
$\Theta^1$ and $\Theta^2$, respectively, i.e.\ 
\bea
&&
\bar{\Gamma} \kappa^{1m} = \mp \kappa^{1m} \, , ~~~~
\bar{\Gamma} \kappa^{2m} = \pm \kappa^{2m} \, ,
\nonumber\\
&&
\bar{\Gamma} \pi^1 = \mp \pi^1 \, , \hspace{30pt}
\bar{\Gamma} \pi^2 = \pm \pi^2 ~~ 
\cdots\cdots ~~ \hbox{type IIA}, 
\\
&&
\bar{\Gamma} \kappa^{1m} = \mp \kappa^{1m} \, , ~~~~
\bar{\Gamma} \kappa^{2m} = \mp \kappa^{2m} \, ,
\nonumber\\
&&
\bar{\Gamma} \pi^1 = \mp \pi^1 \, , \hspace{30pt}
\bar{\Gamma} \pi^2 = \mp \pi^2 ~~ 
\cdots\cdots ~~ \hbox{type IIB and type I}. 
\nonumber
\eea
The local parameters $\lambda^{1m}$ and $\lambda^{2m}$ 
are 12-dimensional scalars. 
$\kappa^{1m}$, $\kappa^{2m}$, $\lambda^{1m}$, and $\lambda^{2m}$ 
are vectors in two dimensions and satisfy the conditions:
\bea
&&
\kappa^{1m} = P_+^{mn} \kappa^1_n
~~~~~~
\kappa^{2m} = P_-^{mn} \kappa^2_n
\, , \\
&&
\lambda^{1m} = P_-^{mn} \lambda^1_n
~~~~~~
\lambda^{2m} = P_+^{mn} \lambda^2_n
\, . \nonumber
\eea

The Lagrangian \rf{LagrangianGSstring} also has 
the global symmetries: 
Poincar\'e ISO(10,2) symmetry, internal scale symmetry, 
and supersymmetry. 
The Poincar\'e ISO(10,2) transformation 
and the internal scale transformation are 
\bea
\delta \Theta^1  &=&  
\frac{1}{2} r \Theta^1  + 
\frac{1}{2} \omega_{IJ} \Gamma^{IJ} \Theta^1
\, , \nonumber\\
\delta \Theta^2  &=& 
\frac{1}{2} r \Theta^2  + 
\frac{1}{2} \omega_{IJ} \Gamma^{IJ} \Theta^2
\, , \nonumber\\
\delta \xi^I     &=&  \omega^I{}_J \xi^J  +  a^I
\, , \nonumber\\
\delta \tA_m      &=&  r \tA_m + \sum_{i=1}^{2g} \alpha_i h^{(i)}_m
\, , \label{GlobalPoincareGSstring}\\
\delta \auxA^I    &=&  
- r \auxA^I  +  \omega^I{}_J \auxA^J
\, , \nonumber\\
\delta \tB_m^I    &=&  r \tB_m^I  +  \omega^I{}_J \tB_m^J 
    + \sum_{i=1}^{2g} ( \beta_i^I + \alpha_i \xi^I) h^{(i)}_m
\, , \nonumber\\
\delta \tC        &=&  2 r \tC
\, , \nonumber\\
\delta g_{mn}     &=&  0
\, . \nonumber
\eea
As the same as in the {\UvUa} bosonic model, 
using the global symmetries \rf{GlobalPoincareGSstring}, 
the solution \rf{auxiliaryfieldSolution} is achieved 
from the equations of motion without loss of generality. 
Then, the gauge transformations 
$\delta \Theta^1$ and $\delta \Theta^2$ in \rf{PiSymmetry} give 
the light-cone-like gauge fixing condition \rf{LightConeTheta}. 
The supersymmetry transformation is 
\bea
\delta \Theta^1   &=&  \ep^1
\, , \nonumber\\
\delta \Theta^2   &=&  \ep^2
\, , \nonumber\\
\delta \xi^I      &=&  - i ( \ep^1 \Gamma^I{}_J \Theta^1 + 
                             \ep^2 \Gamma^I{}_J \Theta^2 ) \auxA^J
\, , \nonumber\\
\delta \tA_m      &=&  0
\, , \nonumber\\
\delta \auxA^I    &=&  0
\, , \label{GlobalSusyGSstring}\\
\delta \tB_m^I    &=&  
      2i 
    ( \ep^1 \Gamma^I{}_J \Theta^1 P_{-m}{}^n + 
      \ep^2 \Gamma^I{}_J \Theta^2 P_{+m}{}^n ) \Pi_n^J
\nonumber\\
&& + \, \frac{1}{3} \EP_m{}^n ( 
\, 5 \ep^1 \Gamma^{KI} \Theta^1 
   \Theta^1 \Gamma_{KJ} \prt_n \Theta^1 
 - 5 \ep^2 \Gamma^{KI} \Theta^2 
   \Theta^2 \Gamma_{KJ} \prt_n \Theta^2 
\nonumber\\
&&\phantom{+ \, \frac{1}{3} \EP_m{}^n (}\hspace{-6.8pt}
- \ep^1 \Gamma_{KJ} \Theta^1 
  \Theta^1 \Gamma^{KI} \prt_n \Theta^1 
+ \ep^2 \Gamma_{KJ} \Theta^2 
  \Theta^2 \Gamma^{KI} \prt_n \Theta^2 
) \auxA^J
\nonumber\\
&&- \, i \tA_m ( \ep^1 \Gamma^I{}_J \Theta^1 + 
                 \ep^2 \Gamma^I{}_J \Theta^2 ) \auxA^J
\, , \nonumber\\
\delta \tC        &=&  
\frac{1}{18} \EP^{mn} \Big\{ 
\ep^1 \Gamma^{IJ} ( 
  \Theta^1 \prt_m \Theta^1 \Gamma_{IJ} \prt_n \Theta^1 
+ 2 \prt_m \Theta^1 \prt_n \Theta^1 \Gamma_{IJ} \Theta^1 ) 
\nonumber\\
&&\phantom{\frac{1}{18} \EP^{mn} \Big\{}\hspace{-14.7pt}
- \ep^2 \Gamma^{IJ} ( 
  \Theta^2 \prt_m \Theta^1 \Gamma_{IJ} \prt_n \Theta^2 
+ 2 \prt_m \Theta^2 \prt_n \Theta^2 \Gamma_{IJ} \Theta^2 ) 
\Big\}
\, , \nonumber\\
\delta g_{mn}     &=&  0
\, . \nonumber
\eea
Performing the supersymmetry transformation 
\rf{GlobalSusyGSstring} twice, 
one obtains 
\beq{GSstringTranslation}
\delta \xi^I  \ = \ 
- i ( \ep^1 \Gamma^I{}_J \ep^{\prime 1} + 
      \ep^2 \Gamma^I{}_J \ep^{\prime 2} ) \auxA^J  \, .
\eeq
This means that the background supersymmetry in this model is 
slightly different from 
the standard super-Poincar\'e ISO(10,2) symmetry 
because there is no translation to the direction $\auxA^I$. 

Finally, we summarize the global symmetries in the following. 
\bea
&&\hbox{i)}  \hspace{12pt}  \hbox{
           $\xi^I \rightarrow -\xi^I$, 
           $\auxA^I \rightarrow -\auxA^I$,
           $\tB_m^I \rightarrow -\tB_m^I$,
           otherwise unchanged,}    
\label{DiscreteTransfGSsuper1}\\
&&\hbox{ii)}  \hspace{8pt}  \hbox{
           $\Theta^1 \rightarrow -\Theta^1$, 
           otherwise unchanged,} 
\label{DiscreteTransfGSsuper2}\\
&&\hbox{iii)}  \hspace{5pt}  \hbox{
           $\Theta^2 \rightarrow -\Theta^2$, 
           otherwise unchanged.} 
\label{DiscreteTransfGSsuper3}
\eea
It should be noted that 
there is no counterpart of the discrete symmetry 
\rf{DiscreteTransfBosonic2} or \rf{DiscreteTransfSuper2} 
in the Green-Schwarz type of {\UvUa} superstring. 
This fact means unfortunately that 
it is not so straightforward to understand the equivalence of 
the Neveu-Schwarz-Ramond type and the Green-Schwarz type of 
{\UvUa} superstring. 
Anyway, in order to show the equivalence, 
one needs to perform the quantization of both theories.

\section{Discussions and Conclusions} 
\label{sec:Membrane}

The form of Lagrangian \rf{LagrangianBosonicGravityModified} 
suggests us that 
the Lagrangian \rf{LagrangianBosonicGravityModified} 
will be naturally extended to the Lagrangian of 
higher dimensional object like membrane or $p$-brane. 
For example, 
$\prt_m \auxA^I = 0$, 
$\auxA_I \prt_m \xi^I = 0$ and 
$\auxA^I \auxA_I = 0$ 
in \rf{EOMBosonic} are obtained by 
the compactification on the internal space--time with condition 
$\auxA^I = \prt_{\widehat{+}} \xi^I$, where $\auxA^I$ is constant. 
The $\pi$-symmetry is considered to be the third component of 
the $\kappa$-symmetry. 
Thus, the extension from {\UvUa} string to (2,2)-brane\footnote{
  The world-sheet swept by string is 
  the space--time with metric ($-$,$+$),
  on the other hand, 
  the world-volume swept by (2,2)-brane is 
  the space--time with metric ($-$,$-$,$+$,$+$).
}
is one of the natural ways. 


It is easy to extend the supersymmetry to $N=2$ supersymmetry. 
In the case of 
the $N=2$ {\UvUa} superstring model, 
the total conformal charge is 
\bea
w^{(N=2)}  &=&  
2 \!\times\! (-13) + 4 \!\times\! \frac{11}{2} + 2 \!\times\! (-1)
\label{ConformalWeightSuper2}\\
&&+ \, 
2 \!\times\! (-1) + 4 \!\times\! (-\frac{1}{2}) + 2 \!\times\! (-1) + 
2D \!\times\! (1 + \frac{1}{2}) 
\nonumber\\
&=&  3 ( D - 4 ) 
\, . \nonumber
\eea
The cancellation of superconformal anomaly, i.e.\ 
the super-Weyl symmetry at the quantum level 
requires $w^{(N=2)} = 0$. 
Thus, we obtain $D=4$ from \rf{ConformalWeightSuper2}. 
It should be noted that 
the background space--time has two time coordinates, i.e.\ 
$D = 2 + 2 \neq 3 + 1$. 


The followings are the conclusions. 
We have succeeded in obtaining 
the covariant expression of {\UvUa} models 
proposed in ref.\ \cite{U1U1string}, 
which are bosonic and supersymmetric models 
without and with (super)gravity. 
The {\UvUa} bosonic and supersymmetric models without gravity 
have ISO($D-1$,1) Poincar\'e symmetry ($D \ge 2$). 
The {\UvUa} bosonic and supersymmetric models with gravity, 
i.e.\ the {\UvUa} bosonic string and superstring models 
have ISO(26,2) and ISO(10,2) Poincar\'e symmetry, respectively. 
We also obtain the Green-Schwarz type of {\UvUa} superstring model. 
It is also possible to construct 
the {\UvUa} heterotic superstring by using usual method.
In any cases 
the generalized Chern-Simons term plays an important role 
to covariantize the Lagrangian. 

The form of covariantized Lagrangian 
in the {\UvUa} (super)string models 
suggests 
that these models are defined naturally by 
more than two-dimensional field theories, 
namely, 
that membrane or $p$-brane is more fundamental 
than string in these models. 
The {\UvUa} string models are the first examples which 
suggest higher dimensional object 
like membrane or $p$-brane 
in the framework of perturbative field theory 
without using the concept of \lq\lq string duality". 

The relation between the {\UvUa} superstring model, 
M-theory, and F-theory is unfortunately still unclear. 
Though M-theory is a kind of membrane theory, 
it does not have ISO(10,2) Poincar\'e symmetry. 
So, the {\UvUa} superstring model will be directly related with 
F-theory which has ISO(10,2) Poincar\'e symmetry 
and is considered to be based on (2,2)-brane. 


\vspace{36pt}
\noindent
{\Large \bf Acknowledgements}
\vspace{12pt}

We would like to thank 
T.\ Kamimura, T.\ Tsukioka and J.\ Ambj\o rn 
for useful discussions. 
We acknowledges 
the support of the Professor Jan Ambj\o rn 
and the hospitality at the Niels Bohr Institute, 
where part of this work was done.



\vspace{36pt}
\noindent
{\Large \bf Appendix A~~ Notations}
\vspace{12pt}

The two-dimensional space--time indices $m$, $n$ run $0$ and $1$. 
The two-dimensional flat metric and the anti-symmetric tensor are 
\beq{metric}
\eta_{mn} \ = \ 
\eta^{mn} \ = \ 
\left( \begin{array}{cc}
          -1 & 0 \\
           0 & 1
       \end{array}
\right) \, ,  \qquad
\ep^{mn} \ = \  
\left( \begin{array}{cc}
           0 & -1 \\
           1 &  0
       \end{array}
\right) \, .
\eeq
In the curved two-dimensional space--time, 
the metric is $g_{mn}$ and 
the anti-symmetric tensor is $\EP^{mn} = \ep^{mn}/ \sqrt{-g}$, 
where $g = \det g_{mn}$. 
The decomposition of vectors to 
self-dual and anti-self-dual pieces 
is achieved using the projection tensors 
\beq{ProjectionOperator}
P_\pm^{mn}  \ = \  \frac{g^{mn} \pm \EP^{mn}}{2}   \, ,
\eeq
which satisfy the projection conditions, 
$P_\pm^m{}_k P_\pm^{kn} = P_\pm^{mn}$ and
$P_\pm^m{}_k P_\mp^{kn} = 0$. 
The relation 
$P_\pm^{kl} P_\pm^{mn} = P_\pm^{kn} P_\pm^{ml}$ 
is also useful. 
The covariant derivative $\nabla\!_m$ operates to fields as 
\bea
\nabla\!_m \phi  &=&  \prt_m \phi
\, , \nonumber\\
\nabla\!_m A_n  &=&  \prt_m A_n - \Gamma^l{}_{mn} A_l
\, , \\
\nabla\!_m A^n  &=&  \prt_m A^n + \Gamma^n{}_{ml} A^l
\, , \nonumber
\eea
where 
$\Gamma^l{}_{mn} = 
 \half g^{lk} (\prt_m g_{kn} + \prt_n g_{mk} - \prt_k g_{mn})$.

The spinor metric is 
\beq{spinormetric}
\eta_{\a\b} \ = \ 
\eta^{\a\b} \ = \ 
\left( \begin{array}{cc}
           0 & 1 \\
          -1 & 0
       \end{array}
\right) \, .
\eeq
The spinor indices are raised or lowered as 
\beq{spinorupanddown}
\th^\a = \eta^{\a\b} \th_\b \, , \quad
\th_\a = \th^\b \eta_{\b\a}
\, .
\eeq
The $\sg$-matrices satisfy
\beq{gammamatrix}
\{ \sg^m , \sg^n \} \ = \ 2 \eta^{mn}
\, .
\eeq
The explicit expression of $\sg$-matrices is 
\beq{gammamatrix01}
(\sg^0)_\a{}^\b \ = \ 
\left( \begin{array}{cc}
           0 & 1 \\
          -1 & 0
       \end{array}
\right) \, , \qquad
(\sg^1)_\a{}^\b \ = \ 
\left( \begin{array}{cc}
           0 & 1 \\
           1 & 0
       \end{array}
\right) \, ,
\eeq
and 
\beq{gammamatrix5}
(\bar{\sg})_\a{}^\b \ = \ 
(\sg^0\sg^1)_\a{}^\b \ = \ 
\left( \begin{array}{cc}
           1 & 0 \\
           0 & -1
       \end{array}
\right) \, .
\eeq
The inner-product of spinors is defined by 
\beq{spinorinnerproduct}
\th {\cal M}_\sg \chi \ = \ 
\th^\a ({\cal M}_\sg)_\a{}^\b \chi_\b
\, ,
\eeq
where ${\cal M}_\sg$ represents any product of $\sg$-matrices. 
The integration of spinor coordinates is 
\beq{thetaintegration}
\int\! \dd^2 \th \ = \ 
\half\int\! \dd \th^1 \dd \th^2
\, .
\eeq
The flat super-covariantderivative is 
\beq{flatsupercovariantderivative}
\bDD_\a \ = \ 
\frac{\prt}{\prt \th^\a} + i ( \sg^m \th )_\a \prt_m
\, .
\eeq
In the curved super-space--time, 
the super-covariantderivative 
${\cal D}_\a = E_\a{}^M {\cal D}_M$ 
operates to superfields as 
\bea
{\cal D} \Phi    &=&  \dd \Phi
\, , \nonumber\\
{\cal D} \Psi\!{}_A  &=&  \dd \Psi\!{}_A + \Omega \, \ep_A{}^B \Psi_B
\, , \\
{\cal D} \Psi^A  &=&  \dd \Psi^A - \Omega \, \Psi^B \ep_B{}^A
\, , \nonumber
\eea
where 
${\cal D} = \dd z^M {\cal D}_M$, 
$\dd = \dd z^M \prt_M$ and $\Omega = \dd z^M \Omega_M$. 

The $D$-dimensional space--time indices $I$, $J$ run 
$1$, $\ldots$, $D-2$, $\widehat{0}$, $\widehat{1}$ 
in the case of {\UvUa} (supersymmetric) models without gravity, 
and 
$0$, $1$, $\ldots$, $D-3$, $\widehat{0}$, $\widehat{1}$ 
in the case of {\UvUa} (supersymmetric) models coupled to gravity. 
The {\UvUa} (supersymmetric) models coupled to gravity are 
equivalent to the {\UvUa} (super)string models.
$D$-dimensional space--time flat metric is 
\beq{backgroundmetric}
\eta_{IJ} \ = \ 
\eta^{IJ} \ = \ 
{\rm diag} ( \underbrace{1, \ldots, 1}_{D-2}, -1, 1 )
\, .
\eeq
in the case of {\UvUa} (supersymmetric) models without gravity, 
and 
\beq{backgroundmetricGravity}
\eta_{IJ} \ = \ 
\eta^{IJ} \ = \ 
{\rm diag} ( -1, \underbrace{1, \ldots, 1}_{D-3}, -1, 1 )
\, .
\eeq
in the case of {\UvUa} (super)string models. 
The value of $D$ is 
$D \ge 2$ for {\UvUa} (supersymmetric) models without gravity, 
$D=28$ for {\UvUa} bosonic string model, and 
$D=12$ for {\UvUa} superstring model. 

The 12-dimensional $\Gamma$-matrices are 
$64 \!\times\! 64$ matrices and satisfy
\beq{GammaMatrix12D}
\{ \Gamma^I , \Gamma^J \} \ = \ 2 \eta^{IJ}
\, ,
~~~~~~~~
(\Gamma^I)^\dagger  \ = \  \Gamma_I
\, .
\eeq
The explicit expression of $\Gamma$-matrices is 
\bea\label{GammaMatrix12D01}
&&
\Gamma^{\widehat{0}}  \ = \ 
\left( \begin{array}{cc}
           0 & 1 \\
          -1 & 0
       \end{array}
\right) \, , \qquad
\Gamma^{\widehat{1}}  \ = \ 
\left( \begin{array}{cc}
              0     & \bar{\g} \\
           \bar{\g} &    0
       \end{array}
\right) \, , 
\\
&&
\Gamma^0  \ = \ 
\left( \begin{array}{cc}
              0   & \g^0 \\
            \g^0  &    0
       \end{array}
\right) \, , \hspace{21.9pt}
\Gamma^i  \ = \ 
\left( \begin{array}{cc}
              0  & \g^i \\
           \g^i  &    0
       \end{array}
\right) \, ,
\nonumber
\eea
and 
\beq{GammaMatrix12D5}
\bar{\Gamma}  \ = \ 
\Gamma^{\widehat{0}} \Gamma^{\widehat{1}} 
\Gamma^0 \Gamma^1 \cdots \Gamma^9  \ = \ 
\left( \begin{array}{cc}
           1 & 0 \\
           0 & -1
       \end{array}
\right) \, , \qquad
\bar{\g}  \ = \  \g^0 \g^1 \cdots \g^9  \, ,
\eeq
where $\g^0$, $\g^i$, $\bar{\g}$ are 
10-dimensional $32 \!\times\! 32$ $\g$-matrices which satisfy 
$(\g^0)^2 = -1$, $\{ \g^0 , \g^i \} = 0$, 
$\{ \g^i , \g^j \} = 2 \delta^{ij}$. 
The inner-product of 12-dimensional spinors is defined by 
\beq{spinorinnerproduct12D}
\Theta {\cal M}_\Gamma \Psi \ = \ 
\Theta^\dagger \Gamma^{\widehat{0}} \Gamma^0 {\cal M}_\Gamma \Psi
\, ,
\eeq
where ${\cal M}_\Gamma$ represents any product of $\Gamma$-matrices. 
If we choose the expression of $\Gamma$-matrices which satisfy 
$(\Gamma^I)^\ast = \Gamma^I$, 
then the Majorana condition for a spinor $\Psi$ 
becomes the simplest form, $\Psi^\ast = \Psi$. 
The following relations are useful 
to check the local gauge symmetries of 
the Green-Schwarz type of {\UvUa} superstring model:
\bea
&&
\Gamma^K \Gamma^{IJ}  \ = \ 
\Gamma^{KIJ} + \eta^{KI} \Gamma^J - \eta^{KJ} \Gamma^I
\, ,
\\
&&
\Theta \Gamma^{I_1 \ldots I_n} \Psi  \ = \ 
\Psi   \Gamma^{I_n \ldots I_1} \Theta
\, ,
\\
&&
\Theta \Gamma^{KI} ( 
\Psi_1 \Psi_2 \Gamma_K{}^J \Psi_3 + 
\Psi_2 \Psi_3 \Gamma_K{}^J \Psi_1 + 
\Psi_3 \Psi_1 \Gamma_K{}^J \Psi_2 )
\, + \, \{ I \leftrightarrow J \}
\\
&&~~~
= \ 
\frac{1}{16} \eta^{IJ} \Big( 
7 \Theta \Gamma^{KL} \Psi_1 \Psi_2 \Gamma_{KL} \Psi_3 
- \frac{1}{6!} \Theta \Gamma^{KLMNPQ} \Psi_1 \Psi_2 \Gamma_{KLMNPQ} \Psi_3 
\Big)
\, , \nonumber
\eea
where $\Gamma^{IJ\cdots}$ are defined by 
\bea
\Gamma^{IJ}  &=&  \half ( \Gamma^I \Gamma^J - \Gamma^J \Gamma^I )
\, ,\label{GammaMultiproducts}\\
\Gamma^{IJK} &=&  \frac{1}{6!} ( 
  \Gamma^I \Gamma^J \Gamma^K - \Gamma^J \Gamma^I \Gamma^K
+ \{ \hbox{cyclic permutation} \} )
\, ,
\eea
and so on.

\end{document}